\documentclass[a4paper]{aa}

\usepackage{psfig}
\usepackage{graphicx}
\usepackage{natbib}
\usepackage[english]{babel}  
\usepackage{txfonts}

\begin{document}

\title{Coronal properties of G-type stars \\
in different evolutionary phases}

\author{L. Scelsi\inst{1} \and A. Maggio\inst{2} \and G. Peres\inst{1}  \and R. Pallavicini\inst{2}}

\offprints{L. Scelsi, \email{scelsi@oapa.astropa.unipa.it}}

\institute{Dipartimento di Scienze Fisiche ed Astronomiche, Sezione di Astronomia, Universit\`a di Palermo, Piazza del Parlamento 1, 90134 Palermo, Italy
\and 
INAF - Osservatorio Astronomico di Palermo, Piazza del Parlamento 1, 90134 Palermo, Italy
}

\date{Received, accepted}

\authorrunning{L. Scelsi et al.}
\titlerunning{Coronal properties of G-type stars}

\abstract{We report on the analysis of XMM-\emph{Newton} observations of three
G-type stars in very different evolutionary phases: the weak-lined T Tauri star
\object{HD 283572}, the Zero Age Main Sequence star \object{EK Dra} and the
Hertzsprung-gap giant star \object{31 Com}. They all have high X-ray luminosity 
($\sim 10^{31}$\,erg\,s$^{-1}$ for HD 283572 and 31 Com and  
$\sim 10^{30}$\,erg\,s$^{-1}$ for EK Dra). We compare the Emission Measure
Distributions ($EMD$s) of these active coronal sources, derived from
high-resolution XMM-\emph{Newton} grating spectra, as well as the pattern of
elemental abundances vs. First Ionization Potential (FIP). We also perform
time-resolved spectroscopy of a flare detected by XMM from EK Dra. We interpret
the observed $EMD$s as the result of the emission of ensembles of magnetically
confined loop-like structures with different apex temperatures. Our analysis
indicates that the coronae of HD 283572 and 31 Com are very similar in terms of
dominant coronal magnetic structures, in spite of differences in the
evolutionary phase, surface gravity and metallicity. In the case of EK Dra the
distribution appears to be slightly flatter than in the previous two cases,
although the peak temperature is similar.

}

\maketitle

\section{Introduction}

During the last decade, the analysis of high-resolution X-ray spectra of 
late-type stars, obtained with EUVE, XMM-\emph{Newton} and \emph{Chandra}
\citep[e.g.][]{MonsignoriFossi1995,Schmitt1996,Guedel_EK1997,GriffithsJordan1998,Laming1999,Sanz2002,Argy2003}, revealed that the thermal structure of coronal
plasmas is better described by a continous Emission Measure Distribution,
$EMD$, rather than by the combination of a few isothermal components usually
employed to fit low- and medium-resolution spectra. Since the coronal plasma is
optically thin, the $EMD$ of the whole stellar corona can be viewed as the sum
of the emission measure distributions of all the loop-like structures where the
plasma is magnetically confined; therefore, it can be used to derive information
about the properties of the coronal structures and the loop populations
\citep{Peres2001}. In particular, the studies mentioned above have indicated
that the coronae of intermediate and high activity stars appear to be more
isothermal than coronae of solar-type stars, and that the bulk of the plasma
emission measure is around $\log T \sim 6.6$ for stars of intermediate activity
and up to $\log T \sim 7$ for very active stars. The latter result is consistent
with the one previously obtained from the analyses of \emph{Einstein} and
ROSAT data, i.e. that there is a good correlation between the effective
coronal temperature and the X-ray emission level \citep[see, for example, ][]{Schmitt1990,Preibisch1997}.

The observation that in active stars a considerable amount of plasma steadily
resides at very high temperatures, which are achieved on the Sun only during
flaring events, led to the hypothesis that a superposition of unresolved flares
may heat the plasma causing an enhanced quasi-quiescent coronal emission level.
Following this idea, \citet{Guedel1997} showed that the time-averaged
$EMD$ resulting from hydrodynamic simulations of a statistical set of flares,
distributed in total energy as a power law, could be made quite similar to the
$EMD$ of stars of different activity level (and age). In particular, he
obtained distributions with two peaks and a minimum around 10\,MK; the
amount of the hottest plasma (at $\sim 12-30$\,MK) decreases with decreasing
$L_{\rm X}$ (or, equivalently, with increasing age) and, at the same time, the
first peak moves towards lower temperatures.

A qualitatively different scenario for the evolution of the $EMD$ with
activity has been proposed by J. Drake  
\citep[see Fig. 2 in the review by ][]{Bowyer2000}: the distribution increases
monotonically from the minimum, which occurs at $\log T$ between $\sim 5$ and
$6$, up to the peak at coronal temperatures; the location of the 
peak shifts towards higher and higher temperatures (up to $\log T \sim 7$ in the
most active stars) for increasing X-ray activity level. Along with the shift of
the peak, the steepness of the ascending part of the distribution increases.

In the pictures sketched above, the shape of the $EMD$ changes with the
stellar activity level; however, it is not yet understood which stellar
parameters (luminosity, surface flux, surface gravity, evolutionary phase, or
others) have a major role in determining the physical characteristics of
the dominant coronal structures and, hence, the properties of the whole Emission
Measure Distribution.

In order to investigate this issue, we have examined the cases of three G-type
stars, in different evolutionary phases: the Pre-Main Sequence star HD 283572,
the Zero-Age Main Sequence star EK Draconis (HD 129333) and the Hertzsprung-gap
giant star 31 Com (HD 111812). Here we report on the analyses of recent
XMM-\emph{Newton} observations of these bright targets, characterized by similar
and relatively high X-ray luminosities ($L_{\rm X} \sim 10^{30}$\,erg\,s$^{-1}$
for EK Dra, and $L_{\rm X} \sim 10^{31}$\,erg\,s$^{-1}$ for HD 283572 and 31
Com) with respect to the \object{Sun}. Previous analyses 
\citep[e.g. ][]{Guedel_EK1997,Ayres1998,Favata1998} showed that the
characteristic coronal temperatures of the stars of our sample lie around
$10^{7}$\,K; the EPIC and RGS detectors on board XMM are very sensitive
to this temperature regime, allowing us to get rather accurate and reliable
information about the plasma Emission Measure Distributions of these stars.

The analysis of the XMM observation of 31 Com was reported in
\citet{Scelsi2004}, while the reconstruction of the $EMD$ of HD 283572 using a
high-resolution spectrum is presented here for the first time. For ease of
comparison with these two sources, we have also re-analyzed the XMM observation
of EK Dra using the same method employed for 31 Com and HD 283572, thus ensuring
homogeneity of the results; note however that independent analyses of the same
XMM observation of EK Dra have been published since 2002
\citep[e.g.][]{Guedel2002_b,Telleschi2003} and more recently and comprehensively
by \citet{Telleschi2004} in the context of a study of solar analogs at different
ages. The latter work is complementary to our present study because it considers
stars having similar mass, size and gravity, but largely different $L_{\rm X}$
and coronal temperature.

This paper is organized as follows: we describe the three targets in 
Sect. \ref{Sample} and we present the observations in Sect. \ref{Observations}.
In Sect. \ref{Data_analysis} we describe the data reduction and the methods used
for the analyses of EPIC and RGS spectra. The results are shown in 
Sect. \ref{Results} and discussed in Sect. \ref{Discussion}.  

\section{The sample}
\label{Sample}

\begin{figure}[t]
\begin{center}
\scalebox{0.5}{
\includegraphics{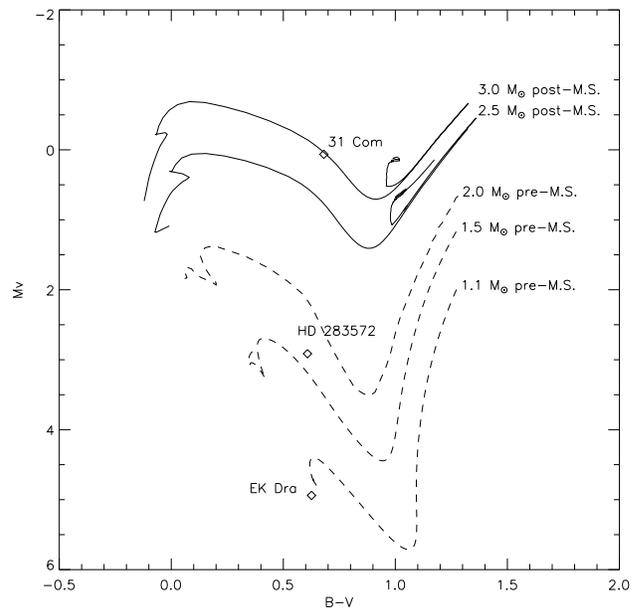}}
\caption{Positions of HD 283572, EK Dra and 31 Com in the H-R diagram. We have
superimposed pre-main-sequence (dashed lines) and post-main-sequence (solid
lines) tracks for the mass values reported in the plot. The evolutionary models
are those of \citet{Ventura1998a,Ventura1998b}, except for the 2\,M$_{\odot}$
pre-M.S. track, for which we have used the model of \citet{Siess2000}. All 
tracks are calculated for solar photospheric abundances.} 
\label{fig:HR}
\end{center}
\end{figure}

In Fig. \ref{fig:HR} we plot the positions of the sample stars in the H-R
diagram, to show their respective evolutionary phases. We used visual
magnitudes, $B-V$ color indexes and distances measured by $Hipparcos$; we
assumed negligible optical extinction in the cases of EK Dra and 31 Com,
coherent with the low interstellar absorption used in the analysis of their
X-ray spectra (Sect. \ref{Results}),
while we used a visual extinction $A_V=0.57$ \citep{Strom1989} and 
$E_{B-V}\sim A_V/3$ for HD 283572.
\begin{center}
\begin{table*}[!t]
\caption{Stellar parameters. Distances are measured by $Hipparcos$; 
$L_{\rm x}$ ($0.3-8$\,keV) are derived in this work from $3-T$ models; 
gravities are determined from the corresponding $M$ and $R$, and surface fluxes
from the corresponding $L_{\rm X}$ and $R$. In the last column, the references
for $L_{\rm bol}$ are indicated in the entries.}
\begin{center}
\begin{tabular}{lcccccccccc} \hline\hline
           & $M/M_{\odot}$ & $R/R_{\odot}$ & Spectral & $P_{\rm rot}$ & $v \sin i$ &  $d$ & $L_{\rm x}$  & $g/g_{\odot}$ & $F_{\rm x}$ & $L_{\rm x}/L_{\rm bol}$ \\
	   &               &               &   type   &       [d]     &    [Km\,s$^{-1}$]  &   [pc] & [$10^{30}$\,erg\,s$^{-1}$] &        & [$10^{6}$\,erg\,s$^{-1}$\,cm$^{-2}$] &                 \\ \hline
HD 283572  &      1.8$\,^a$ & 2.7$\,^b$; 4.1$\,^a$ &  G2   &   1.55$\,^a$    &   78$\,^a$   &   128  &     $\sim 9$     &  0.25; 0.12  &  20; 9      & $5\times\,10^{-4}\,^c$  \\
EK Dra     &      1.1$\,^d$ & 0.95$\,^d$ &  G1.5V   & 2.75$\,^d$ & 17.3$\,^e$ &  34  &  $\sim 1$  &  1.2$\,^f$   &  18    & $3\times\,10^{-4}\,^g$    \\
31 Com     &      3$\,^h$ & 9.3$\,^h$ &  G0III   & $<7.2\,^i$    & 66.5$\,^j$ &  94  &  $\sim 7$  &     0.035    &  1.3   & $3\times\,10^{-5}\,^h$    \\ \hline
\multicolumn{11}{l}{\footnotesize{$^a$ \citet{Strassmeier1998}.}} \\
\multicolumn{11}{l}{\footnotesize{$^b$ \citet{Walter1987} and the $Hipparcos$ measurement of $d$.}} \\
\multicolumn{11}{l}{\footnotesize{$^c$ \citet{Walter1988} and the $Hipparcos$ measurement of $d$.}} \\
\multicolumn{11}{l}{\footnotesize{$^d$ \citet{Guinan2003}.}} \\
\multicolumn{11}{l}{\footnotesize{$^e$ \citet{Strassmeier1998b}.}} \\
\multicolumn{11}{l}{\footnotesize{$^f$ Also consistent with the estimate by \citet{Strassmeier1998b}.}} \\
\multicolumn{11}{l}{\footnotesize{$^g$ \citet{Redfield2003}.}} \\
\multicolumn{11}{l}{\footnotesize{$^h$ \citet{Pizzolato2000}.}} \\
\multicolumn{11}{l}{\footnotesize{$^i$ From $P_{\rm rot}$ and $v \sin i$.}} \\
\multicolumn{11}{l}{\footnotesize{$^j$ \citet{deMedeiros1999}.}} \\
\end{tabular}
\end{center}
\label{tab:param}
\end{table*}
\end{center}

The latter star is a member of the Taurus-Auriga star forming region and its age
is estimated to be $\sim 2\times 10^6$\,yr \citep{Walter1988}. HD 283572 shows
no sign of accretion from a circumstellar disk, which characterizes the
earlier stage of classical T Tauri stars; the decoupling from the disk allowed
this star to spin up cosiderably, due to its contraction, up to several tens of
km\,s$^{-1}$ ($v\sin i = 78$\,km\,s$^{-1}$, see Table \ref{tab:param}), probably with a
consequently enhanced dynamo action and a very high X-ray luminosity 
($L_{\rm X}\sim 10^{31}$\,erg\,s$^{-1}$). From Fig. \ref{fig:HR} we deduce that 
HD 283572 will be an A-type star during its main sequence phase, and we estimate
a mass between $\sim 1.5$ and $\sim 2$\,M$_{\odot}$, in agreement with the
estimate of $1.8\pm 0.2$\,M$_{\odot}$ by \citet{Strassmeier1998}. The radius of
HD 283572 has been derived by \citet{Walter1987} through the Barnes-Evans
relation, $R \sim 3.3$\,R$_{\odot}$ at an assumed distance of 160\,pc, which
becomes 2.7\,R$_{\odot}$ at the new distance of 128\,pc measured by $Hipparcos$;
more recently, \citet{Strassmeier1998} combined photometric measurements,
rotational broadening and Doppler imaging technique to determine the radius of
HD 283572 in the range $3.1-4.7$\,R$_{\odot}$, with a best value of
4.1\,R$_{\odot}$. Due to the uncertainties of these estimates, we decided to
consider both of them. We anticipate that our main results are only weakly 
affected by the choice of one of these values.

EK Dra is a G1.5-type star with mass and radius about equal to the solar values.
It has just arrived on the main sequence, thus representing an analog of the
young Sun. Because of its age 
\citep[$\sim 7\times 10^7$\,yr, ][]{Soderblom1987}, it suffered little magnetic
braking and its short rotational period \citep[$\sim 2.7$ days, ][]{Guinan2003}
makes it a bright X-ray source ($L_{\rm X}\sim 10^{30}$\,erg\,s$^{-1}$). 

The more massive ($M\sim 3$\,M$_{\odot}$) giant star 31 Com 
\citep[age $\sim 4\times 10^8$\,yr, ][]{Friel1992} has already evolved out of
the main sequence and now it is crossing the Hertzsprung-gap. The position of 
31 Com in the H-R diagram and the evolutionary models indicate a spectral type
late-B/early-A on the main sequence; therefore, this star has developed a
convective subphotospheric layer and a dynamo only in its current post-main 
sequence evolutionary phase \citep{Pizzolato2000}. The X-ray luminosity is 
$\sim 7\times 10^{30}$\,erg\,s$^{-1}$.

The stellar parameters of the three targets, with the relevant references, are
summarized in Table \ref{tab:param}. For HD 283572 we report both estimates,
mentioned above, of the stellar radius and the corresponding values of gravity
and surface X-ray flux.

HD 283572, EK Dra and 31 Com  were chosen because their stellar parameters
offer the possibility to get useful insight into their coronal properties from the
comparison of their $EMD$. Note, in particular, that while the X-ray
luminosity of 31 Com and HD 283572 are about equal and larger than that of 
EK Dra by about an order of magnitude, EK Dra and HD 283572 are the stars with
the highest surface fluxes, whose values exceed significantly that of 31 Com,
by about one order of magnitude. Note also that the different evolutionary
phases imply different stellar internal structures; moreover, these targets have
quite different gravities, implying different pressure scale heights and
possible changes in the properties of the dominant coronal loops. 

Finally, the rapidly rotating stars HD 283572 and 31 Com are putative single
sources: this avoids difficulties in the interpretation of the results, due both
to the uncertain origin of the emission, in case of multiple components,
and to the possibility of an enhanced activity as found, for example, in 
tidally-locked RS CVn systems. On the contrary, EK Dra has a distant companion
\citep{Duquennoy1991}, whose mass is likely between 0.37\,M$_{\odot}$ and
0.45\,M$_{\odot}$ \citep{Guedel1995b}. \citet{Guedel1995a} found that the X-ray
and radio emissions are modulated with the rotational period, strongly
suggesting that the coronal emission comes predominantly from the G star.
If we assume that the secondary star has $M \sim 0.4$\,M$_{\odot}$ and
age $\sim 70$\,Myr, and  has a saturated corona (the worst case), its X-ray
luminosity would be $\sim 10^{29}$\,erg\,s$^{-1}$, so we might expect contamination of
the X-ray emission of the G star from the companion at most at $\sim 10$
\% level.

\section{Observations}
\label{Observations}

The observations of HD 283572, EK Dra and 31 Com were performed
with  XMM-\emph{Newton} respectively on September, 5, 2000 (PI: R. Pallavicini),
on December, 30, 2000 (PI: A. Brinkman) and on January 9, 2001 
(PI: Ph. Gondoin). The non-dispersive CCD cameras 
\citep[EPIC MOS and EPIC {\sc pn},][]{Turner2001,Struder2001}, lying in the
focal plane of the X-ray telescopes, have spectral resolution 
$R=E/\Delta E \sim 5-50$ in the range $0.1-10$\ keV, while the two reflection
grating spectrometers \citep[RGS, ][]{denHerder2001} provide resolution 
$R \sim 70-500$ in the wavelength range $5-38$\ \AA\ ($0.32-2.5$\ keV).

For the present study, we considered only the EPIC {\sc pn} and RGS data; in
Table \ref{tab:log_obs} we report details on the instrument configurations and
on the observations.

At the time of these observations, both CCD 7 of RGS1 and CCD 4 of RGS2
were not operating. These CCDs correspond to the spectral regions
containing the He--like triplets of neon and oxygen, respectively. Note also
that the RGS1 spectrum of HD 283572 is entirely missing, due to instrument setup
problems in the early phase of XMM-\emph{Newton} observations; hence we have no
information on the \ion{O}{vii} triplet for this source.
\begin{center}
\begin{table*}[t]
\caption{Log of the XMM-\emph{Newton} observations.}
\begin{center}
\begin{tabular}{lcccccccccccc} \hline\hline
           & \multicolumn{3}{c}{Exposure time (ks)} & EPIC {\sc pn} & \multicolumn{3}{c}{Q.E. Exposure$^{a}$ (ks)} & &\multicolumn{3}{c}{Count-rate$^{b}$ (s$^{-1}$)} \\
\cline{2-4} \cline{6-8} \cline{10-12}  
	   &      {\sc pn}     &     RGS1     &     RGS2    & Mode/Filter &      {\sc pn}     &     RGS1     &     RGS2    & &  {\sc pn}    &    RGS1    &    RGS2     \\ \hline
HD 283572  &      41.1   &       0      &     48.7    & Full Frame/Medium &      41.1   &       0      &     47.4    & & 2.20   &    0       &    0.15     \\
EK Dra     &      46.9   &      51.7    &     50.2    & Large Window/Thick &      38.5   &      44.9    &     43.6    & & 2.20   &    0.16    &    0.22     \\
31 Com     &      33.5   &      41.7    &     40.5    & Full Frame/Thick &      32.2   &      39.6    &     38.5    & & 1.45   &    0.11    &    0.16     \\ \hline
\multicolumn{12}{l}{\footnotesize{$^a$ Exposure time for the analysis of the quiescent emission (Q.E.), i.e. excluding the time intervals affected by proton flares,}} \\
\multicolumn{12}{l}{\footnotesize{ occurred in the cases of HD 283572 and 31 Com, and by the source flare in the case of EK Dra (see Sect. \ref{light_curves}).}} \\
\multicolumn{12}{l}{\footnotesize{$^b$ Mean count-rate in the $1.2-62$\,\AA\ ($0.2-10$\,keV) band for {\sc pn} and in the $5-38$\,\AA\ ($0.32-2.5$\,keV) band for RGS (1st order }} \\
\multicolumn{12}{l}{\footnotesize{ spectrum) relevant to the Q.E. Exposure.}} \\
\end{tabular}
\end{center}
\label{tab:log_obs}
\end{table*}
\end{center}

\section{Data analysis}
\label{Data_analysis}

We used SAS version 5.3.3, together with the calibration files available at the
time of the analysis (June 2002), to reduce the data of HD 283572 and 31 Com;
the data of EK Dra were reduced with SAS version 5.4 and the analysis was
performed in September 2003. We generated all {\sc pn} responses with the SAS
{\sc rmfgen} and {\sc arfgen} tasks. 

Good Time Intervals were selected by excluding those time intervals showing the
presence of presumable proton flares in the background light curve extracted
from CCD 9 of the RGS, following \citet{denHerder2002}: we cut the
intervals where the count-rate exceeds 0.1 cts\,s$^{-1}$ for 31 Com and 1.6
cts\,s$^{-1}$ in the case of HD 283572 (whose observation is contaminated by
high level of background), while we did not exclude any interval in the case of
EK Dra.

In order to obtain X-ray light curves and spectra, we extracted the events from
a circular region ($\sim 50''$ radius) within CCD 4 for HD 283572, while we used
annular regions ($\sim 7.5''-50''$ radii) for 31 Com and EK Dra, because the
relevant data were affected by pile-up. In all cases, background photons were
extracted from the rest of CCD 4, excluding the sources and their out-of-time
events.

\subsection{Light curves}
\label{light_curves}

\begin{figure}[!h]
\begin{center}
\scalebox{0.41}{
\includegraphics{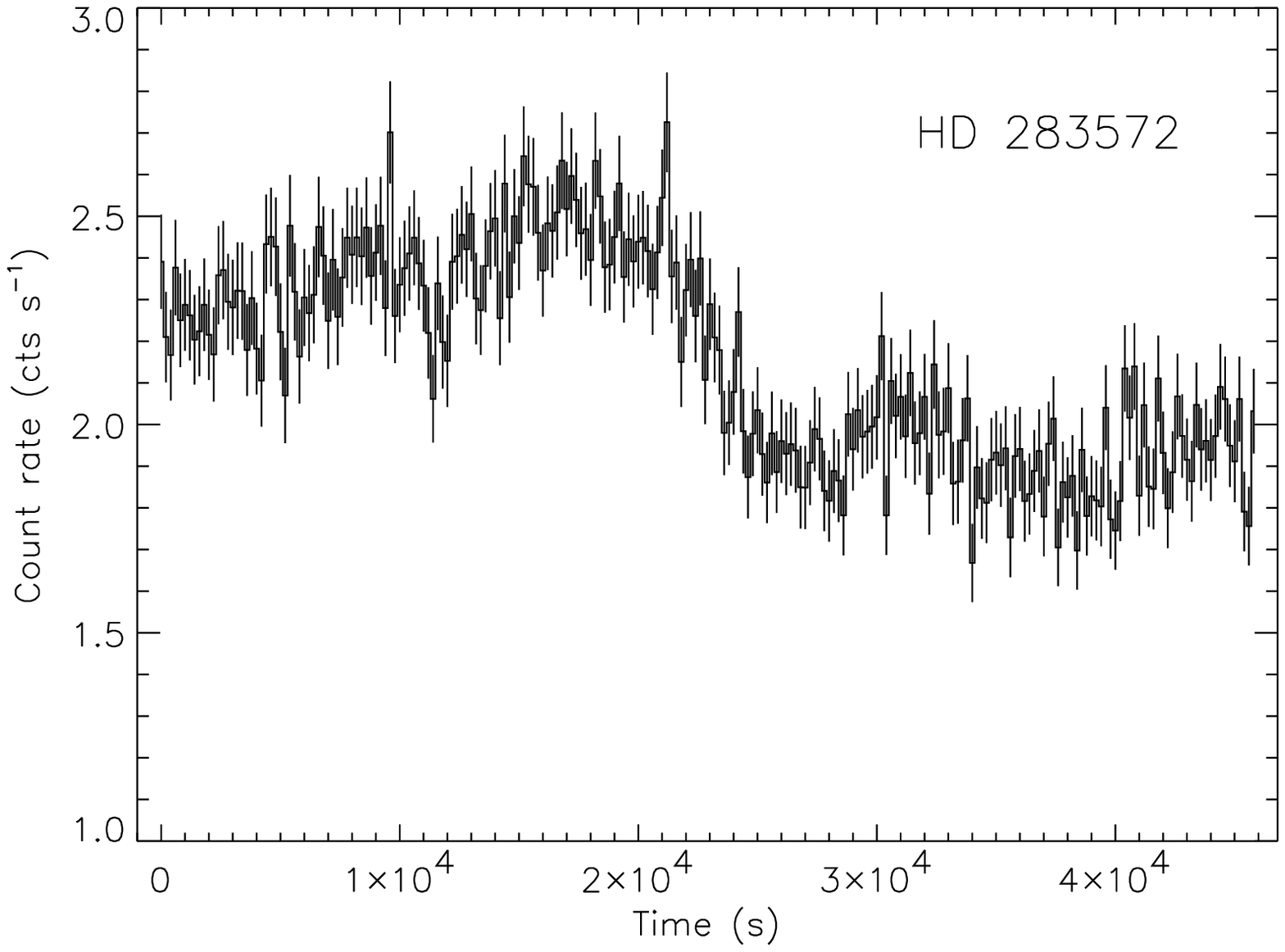}}
\scalebox{0.41}{
\includegraphics{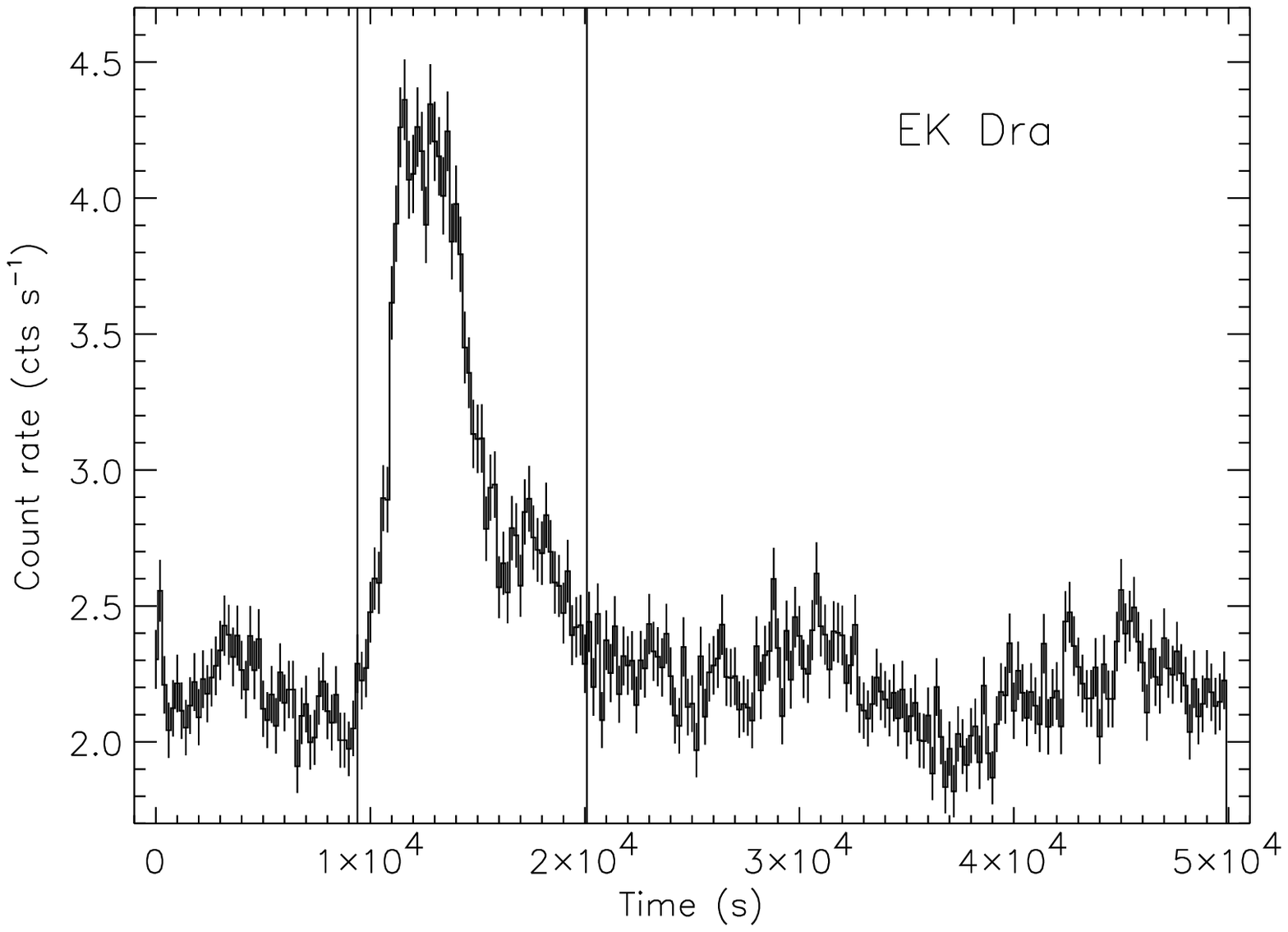}}
\scalebox{0.41}{
\includegraphics{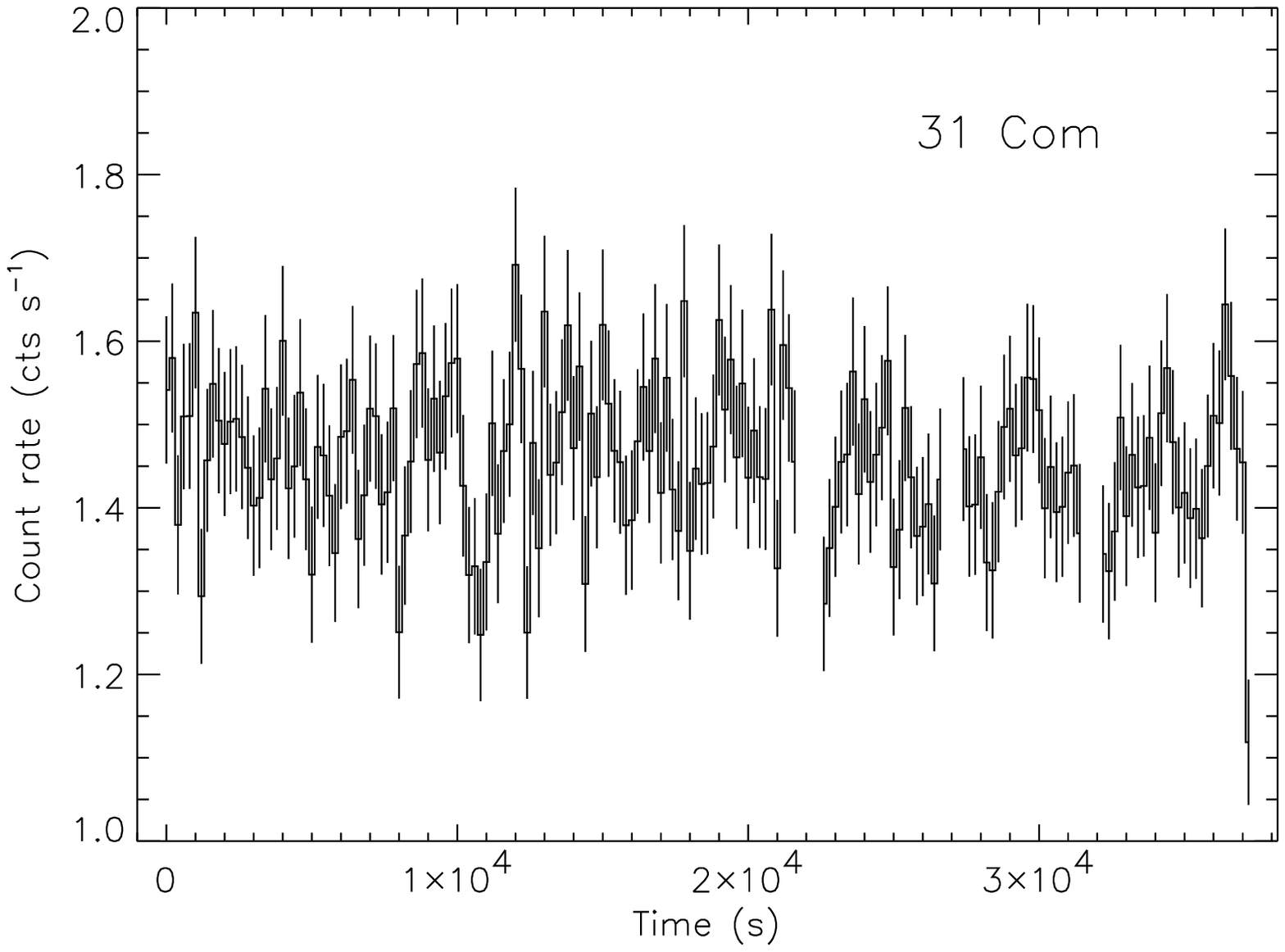}}
\caption{Background-subtracted {\sc pn} light curves of HD 283572 (upper), EK
Dra (middle) and 31 Com (lower), in the $0.2-10$\,keV band and with time bins of
200\,s. The vertical lines in the light curve of EK Dra mark the time interval
of the flare, excluded from the analysis of the quiescent emission.} 
\label{fig:lc}
\end{center}
\end{figure}

Figure \ref{fig:lc} shows the {\sc pn} background-subtracted light curves of the
sources, with a 200\,s time binning. The light curve of 31 Com is the only one
that is consistent with the hypothesis of a constant emission 
\citep[see Sect. 3.1 in ][]{Scelsi2004}.

In the case of HD 283572, there is evidence of variability of the emission, on
a time-scale of the order of 30\,ks, which is not a typical flare
event. The reduced $\chi^2_{\rm r}$ is 5.9 ($229\ d.o.f.$) in the null
hypothesis of a constant emission; the variability amplitude, calculated as
0.5 [max(rate)-min(rate)]/mean(rate), is $\sim 20\%$. There is a less pronounced
variability on a time-scale of $\sim 10$\,ks. From Tables \ref{tab:param} and
\ref{tab:log_obs}, we note that the duration of the observation is about one
third of the stellar rotational period, hence a large fraction of
the stellar surface was visible during the pointing; this suggests that
at least part of the variability is due to an inhomogeneous distributiuon of 
active regions over the stellar surface.

The light curve of EK Dra clearly shows the presence of a flare; the vertical
lines in the figure mark the start and the end of the flare, obtained as the 
minimum and maximum times where the hardness-ratio, 
$HR=(H-S)/(H+S)$\footnote{We have evaluated the soft emission count-rate, $S$,
in the $0.3-1$\,keV band and the hard emission count-rate, $H$, in the
$1-10$\,keV band.}, systematically exceeds by more than 1 $\sigma$ the average
$HR$ value calculated from time intervals before and after the flare. We 
excluded the time interval of the flare from the emission measure analysis
(Sect. \ref{EMR}), since we want to study the thermal properties of the
quiescent corona, and we analyzed the flare separately. The quiescent
emission of EK Dra is still variable, yielding a reduced $\chi^2_{\rm r}=6.2$
($190\ d.o.f.$) against the null hypothesis of a constant source; the
variability is on a time-scale of $\sim 15$\,ks and its amplitude (calculated
as above) is $\sim 16\%$.

\subsection{EPIC {\sc pn} spectra}

\begin{figure*}[!t]
\begin{center}
\scalebox{0.7}{
\includegraphics{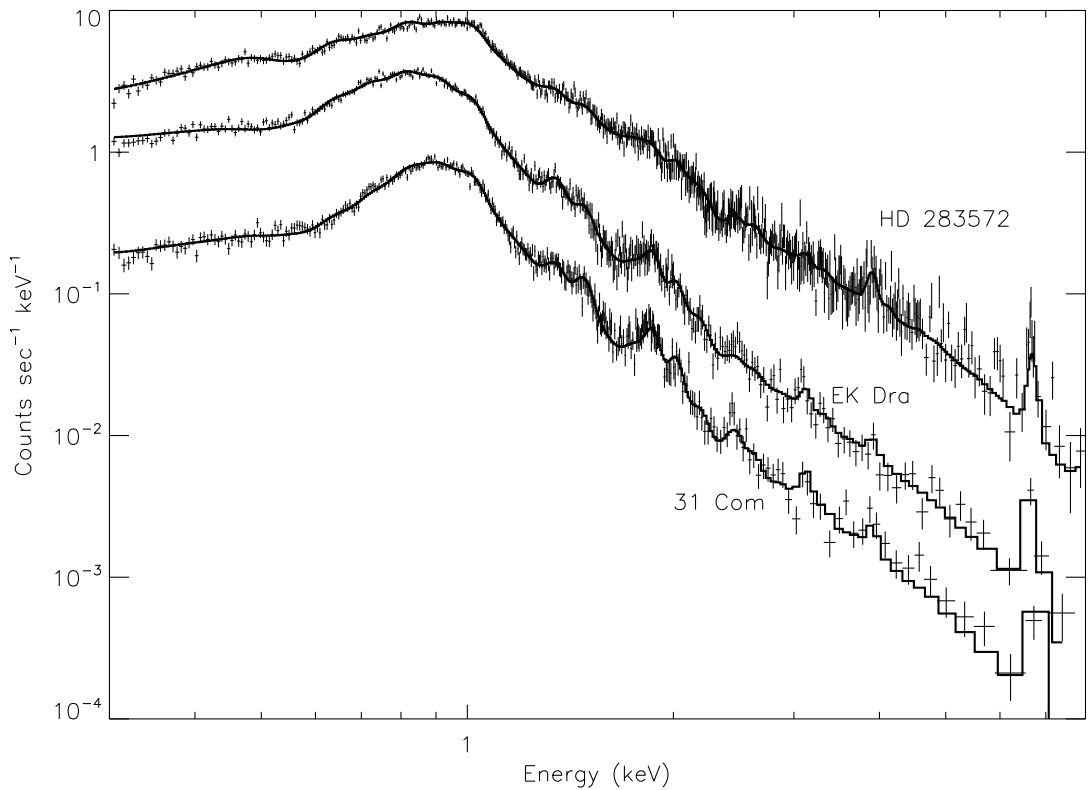}}
\caption{EPIC {\sc pn} spectra of HD 283572, EK Dra and 31 Com with their
best-fit model spectra (the parameters of the models are listed in Table
\ref{tab:fit_EPIC}). The spectra of 31 Com and HD 283572, with their relevant
best-fit models, have been shifted by -0.5 and +0.5 dex for clarity.}
\label{fig:spettri_EPIC}
\end{center}
\end{figure*}
We have performed global fitting of the EPIC {\sc pn} spectra (Fig. 
\ref{fig:spettri_EPIC}) with the aim of deriving, from multi-component thermal
models, the initial guess of the continuum level for the line measurements in
the RGS spectra. Moreover, the abundances of some elements (Si, S, Ar, Ca)
can be better determined from {\sc pn} spectra, rather than from RGS spectra,
thanks to the wider spectral range of the former -- which includes the strong
K-shell lines of the relevant H--like and He--like ions -- and to the higher
photon counting statistics\footnote{The \ion{Si}{xiii-xiv} lines fall also
in the RGS spectral range, but the statistics are usually very low and the
calibration of the effective area is less precise at these wavelengths; 
nonetheless, the results of our analysis show that the Si abundances derived
from {\sc pn} data are consistent with those obtained from RGS spectra within
statistical uncertainties (see Table \ref{tab:abund_RGS}).}.

We analyzed these spectra with XSPEC v11.2 and we found that an absorbed, 
optically-thin plasma with three isothermal components provides an acceptable
description of each of them (see results in Sect. \ref{results3_T}). The models
are based on the Astrophysical Plasma Emission Database (APED/ATOMDB V1.2) and
have variable abundances; we adopted the criterion of leaving free to vary only
the abundances of those elements (O, Ne, Mg, Si, S, Fe, Ni, in some cases Ca and
Ar) with strong and clearly detectable line complexes in EPIC spectra. The
abundances of the other elements were tied to that of iron, their best-fit
values being poorly constrained when left free to vary.

We eventually used the high-energy tail of the {\sc pn} spectrum also to
check the high-temperature tail of the emission measure distributions, as 
described in the next section and in Appendix \ref{check_EK_hot_tail}.

Finally, we performed time-resolved spectroscopy of the {\sc pn} data of EK Dra
during the flare, to get information on the properties (in particular the size)
of the flaring loop, employing the method by \citet{Reale1997}. This analysis
and its results are reported in Appendix \ref{results_flare}.

\subsection{Emission Measure Reconstruction}
\label{EMR}

The approach we adopted for the line-based analysis of the RGS spectra of each
star is discussed in detail in \citet{Scelsi2004} together with a study of its
accuracy; here we limit ourselves to report the main points of our iterative
method. 

We employed the software package PINTofALE \citep{PoA2000} and, in part, also
XSPEC, and used the APED/ATOMDB V1.2 database which includes the
\citet{Mazzotta1998} ionization equilibrium.

We first rebinned and co-added the background-subtracted RGS1 and RGS2 spectra
for the identification of the strongest emission lines and the measurement of
their fluxes. In this latter step, we adopted a Lorentzian line profile 
and we assumed initially the continuum level evaluated from the 3-T model best
fitting the {\sc pn} spectrum, because the wide line wings make it
impossible to determine the true source continuum below $\sim 17$\,\AA\
directly from the RGS data, in particular in the $\sim 10-17$\,\AA\
range, where the spectrum is dominated by many strong overlapping lines.
Then, with the aim to reconstruct the Emission Measure Distribution ($EMD$) vs.
Temperature, we selected a set of lines, among the identified ones, with
reliable flux measurements and theoretical emissivities. Most of them are
blended with other lines, so the measured spectral feature is actually the sum
of the contributions of a number of atomic transitions; accordingly, we
evaluated the ''effective emissivity'' of each line blend as the sum of the
emissivities of the lines which mostly contribute to that spectral feature.
Moreover, we carefully selected only iron lines not blended with lines of other
elements, because the procedure we employed (see below) uses these iron lines in
the first step of the $EMD$ analysis, and estimates of the abundances of the
other elements are not yet available at this step.

We performed the $EMD$ reconstruction with the Markov-Chain Monte Carlo (MCMC)
method by \citet{KashyapDrake1998}. This method yields a volume emission measure
distribution, $EM(T_{k})=dem(T_{k})\,\Delta \log T$, and related statistical
uncertainties $\Delta EM(T_k)$, where 
$dem(T)=n_{\rm e}^{2}\,{\rm d}V/{\rm d} \log T$ is the differential emission
measure of an optically thin plasma and $\Delta \log T =0.1$ is a constant bin
size; the method also provides estimates of element abundances, relative to
iron, with their statistical uncertainties. The iron abundance is estimated by
scaling the emission measure distribution assuming different metallicities
and by comparing the synthetic spectrum with the observed one at 
$\lambda >20$\,\AA\ in the RGS spectrum (this is a spectral region free of
strong overlapping emission lines). Finally, we checked the solution
obtained with the MCMC by comparing (i) the line fluxes predicted from
our solution with the measured ones and (ii) the {\sc pn} model spectrum, based
on the reconstructed $EMD$, with the observed {\sc pn} spectrum at $E > 2$\,keV.
These checks are illustrated respectively in Fig. \ref{fig:check_flx} and in 
Appendix \ref{check_EK_hot_tail}, taking the case of EK Dra as an example
(similar results were obtained for the other two stars). In particular, the
correct prediction of the \ion{O}{vii-viii} line fluxes allowed us to check the
reliability of the amount of plasma in the low-temperature tail of the $EMD$; analogously, the correct prediction of the \ion{Fe}{xxiii-xxiv} line 
fluxes and of the high-energy tail of the {\sc pn} spectrum are important tests
for the reliability of the amount of plasma in the high-temperature tail of the
$EMD$.

We also checked the consistency between the continuum level assumed for flux
measurements and the continuum predicted by the $EMD$. In fact, since our
method is iterative, the continuum assumed for flux measurements in the RGS
range is adjusted at each iteration for consistency with the $EMD$, and it may
become different from that predicted by the $3-T$ model best-fitting the 
{\sc pn} spectrum, which is adopted as initial guess. Therefore, this procedure
ensures that possible cross-calibration offsets between {\sc pn} and RGS do not
affect the final $EMD$.

\section{Results}
\label{Results}

\subsection{3-T models}
\label{results3_T}

\begin{center}
\begin{table*}[!t]
\caption{Best-fit models of the EPIC {\sc pn} data (in the $0.3-8$\,keV band),
with 90\% statistical confidence ranges computed for one interesting parameter
at a time; nominal errors on $T_{i}$ and $EM_{i}$ are at the 10\% level. Element
abundances are relative to the solar ones \citep{Grevesse1992}. Mean
temperatures are calculated as
$<T>=\sum_{i=1}^{3}EM_{i}\,T_{i}/\sum_{i=1}^{3}EM_{i}$. Abundances and hydrogen
column densities without errors were fixed as explained in the text.}
\vspace{0.3cm}
\begin{center}
\begin{tabular}{lccc} \hline\hline
                                    &      HD 283572           &        EK Dra        &       31 Com      \\ \hline 
$\log T_{1,2,3}\,({\rm K})$         &   6.64, 7.04, 7.43       &   6.58, 6.94, 7.33   &  6.44, 6.92, 7.28 \\ 
$\log EM_{1,2,3}\,({\rm cm}^{-3})$  &   53.5, 53.5, 53.7       &   52.5, 52.4, 52.3   &  52.6, 53.1, 53.0 \\ 
$\log <T>\,({\rm K})$               &         7.21             &         6.99         &         7.06      \\ 
   C                                &         0.37             &         0.57         &         0.38      \\ 
   N                                &         0.37             &         0.42         &         1.54      \\ 
   O                                & $0.236\,\pm\,0.014$      & $0.346\,\pm\,0.015$  & $0.58\,\pm\,0.03$ \\ 
   Ne                               & $0.46\,\pm\,0.03$        & $0.83\,\pm\,0.04$    & $2.35\,\pm\,0.14$ \\ 
   Mg                               & $0.32\,\pm\,0.05$        & $0.86\,\pm\,0.06$    & $1.95\,\pm\,0.13$ \\ 
   Si                               & $0.25\,\pm\,0.04$        & $0.59\,\pm\,0.06$    & $1.23\,\pm\,0.11$ \\ 
   S                                & $0.26\,\pm\,0.09$        & $0.15\,\pm\,0.10$    & $0.58\,\pm\,0.20$ \\ 
   Ar                               &         0.37             & $0.82\,\pm\,0.22$    &         1.54      \\ 
   Ca                               & $1.8\,\pm\,0.3$          &         0.83         &         1.54      \\ 
   Fe                               & $0.37\,\pm\,0.01$        & $0.83\,\pm\,0.01$    & $1.54\,\pm\,0.02$ \\ 
   Ni                               & $1.52\,\pm\,0.11$        & $1.80\,\pm\,0.20$    & $4.1\,\pm\,0.3$   \\ 
$N_{\rm H}\,({\rm cm}^{-2})$   & $(8.7\,\pm\,0.4)\times 10^{20}$ &    $3\times 10^{18}$      &     $10^{18}$     \\ \hline
   $\chi^{2}_{\nu} /{\rm d.o.f.}$   &      1.1/688             &      1.26/411        &      1.1/367      \\ \hline 
\end{tabular}
\end{center}
\label{tab:fit_EPIC}
\end{table*}
\end{center}

\begin{figure*}[t!]
\begin{center}
\scalebox{0.9}{
\includegraphics{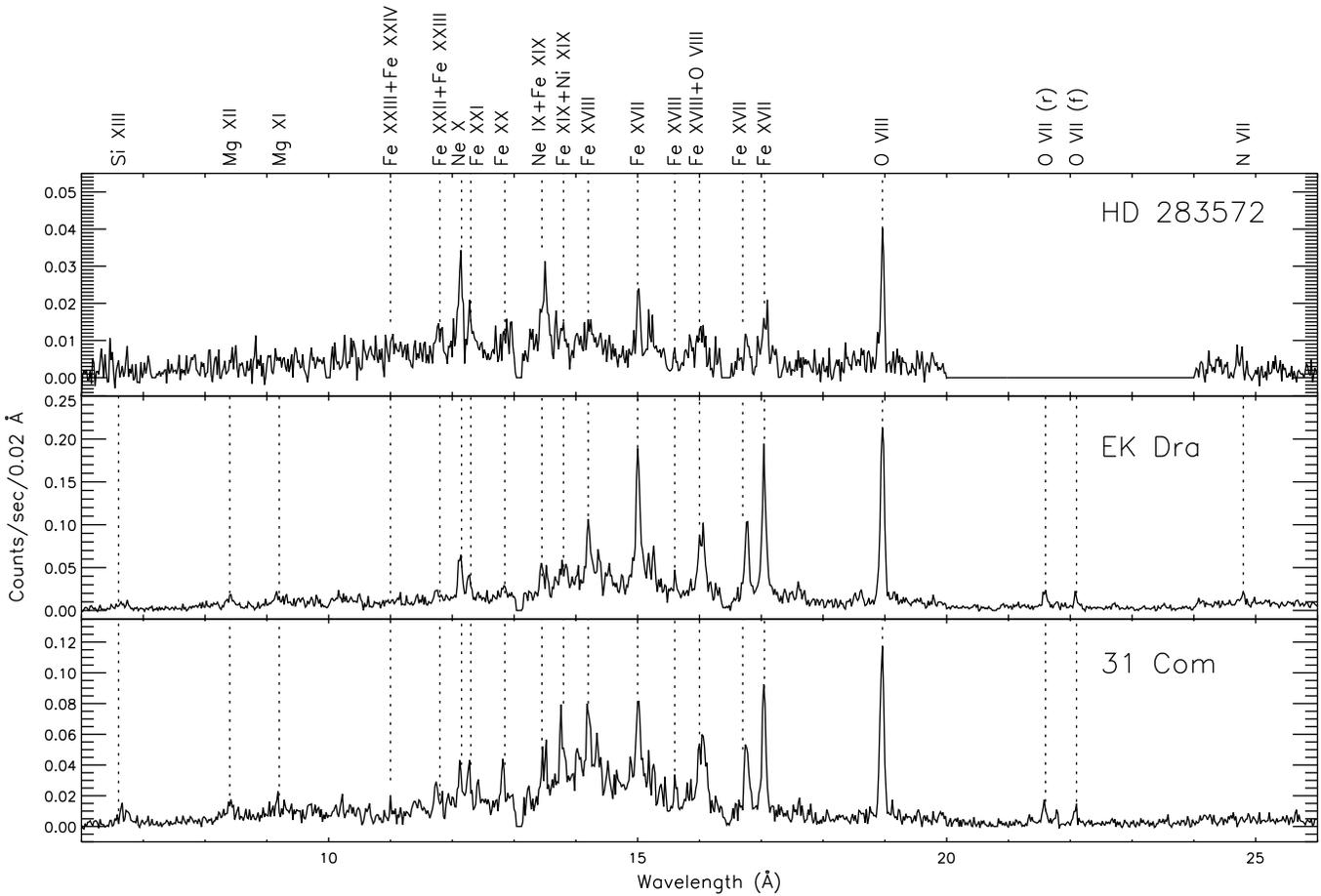}}
\caption{Co-added RGS spectra of HD 283572 (upper), EK Dra (middle) and 31 Com
(lower) with the identification of the most prominent lines; the bin size is
0.02 \AA.}
\label{fig:spettri_RGS_ident}
\end{center}
\end{figure*}

Figure \ref{fig:spettri_EPIC} shows the {\sc pn} spectra with their relevant 
3-T models, obtained by fitting the data in the $0.3-8$\,keV range. The best-fit
parameters of the models are listed in Table \ref{tab:fit_EPIC}.

The presence of the \ion{Fe}{xxv} 6.7 keV emission line in all these spectra is
indicative of hot coronae, as expected from earlier works and confirmed by our
analysis. Note the large best-fit EM values of the hottest components
for all stars, comparable to the EMs of the cooler components; in particular,
the hottest plasma dominates the corona of HD 283572, as also confirmed by the
analysis of $Chandra$ spectra \citep{Audard2004}.

The line complexes of \ion{Mg}{xi-xii} ($\sim 1.3-1.5$\,keV), \ion{Si}{xiii-xiv}
($\sim 1.8-2.1$\,keV) and \ion{S}{xv} ($\sim 2.5$\,keV), as well as the large
bump between 0.6 and 1 keV due to the \ion{Fe}{xvii-xxiv}, \ion{Ni}{xix-xx} and
\ion{Ne}{ix-x} lines allowed us to constrain the abundances of Mg, Si, S, Fe, Ne
and Ni. These complexes are less evident in the spectrum of HD 283572, as a
consequence of the significantly lower metallicity with respect to the other two
stars; instead, the \ion{Ca}{xix} line complex ($\sim 3.9$\,keV) is most
prominent in the spectrum of this star (Fig. \ref{fig:spettri_EPIC}) and
the estimated abundance of this element is higher than for the other two cases.
Note also that we were able to constrain the Ar abundance for EK Dra, thanks to
the clearly visible lines of \ion{Ar}{xvii} at $\sim 3.1$\,keV. In the other
two cases we linked the abundances of Ca and/or Ar to that of Fe assuming
the same ratios as in the solar corona \citep{Grevesse1992}. Moreover, we used
the results of the $EMD$ analyses to fix the coronal C/Fe abundance ratio for 
EK Dra and 31 Com to respectively 0.7 and 0.25 solar, and the N/Fe ratio for
EK Dra to 0.5 solar.

Finally, we could not constrain the interstellar absorption in the 
directions of EK Dra and 31 Com with the fitting procedure, so we fixed them
at the relatively low values of $3\times 10^{18}$\,cm$^{-2}$ and
$10^{18}$\,cm$^{-2}$ measured, respectively, by \citet{Guedel_EK1997} and
\citet{Piskunov1997}. On the contrary, the spectrum of HD 283572 is
significantly absorbed, as expected from the location of this star in the
Taurus-Auriga star forming region. The hydrogen column density we derived
from the fit is compatible with $A_V$ and consistent with previous results
obtained by fitting ASCA, ROSAT, $Einstein$ and SAX data \citep{Favata1998}.

\subsection{Emission Measure Distributions and abundances}
\label{results_EM}

The rebinned and co-added RGS spectra (Fig. \ref{fig:spettri_RGS_ident}) show 
emission lines from \ion{Fe}{xvii-xxiv}, \ion{Ne}{ix-x}, \ion{O}{viii} and
\ion{Ni}{xix-xx} ions in all cases, while \ion{O}{vii}, \ion{Mg}{xi-xii} and
\ion{Si}{xiii-xiv} emission lines are visible only for EK Dra and 31 Com, and
the \ion{N}{vii} line in the case of EK Dra only. Actually, we could not
identify any line outside the wavelength range $10-20$\,\AA\ in the spectrum of
HD 283572, because of contamination from high background and lack of the RGS1
spectrum altogether. The reconstruction of the $EMD$ of EK Dra and 31 Com was
based on about 40 lines, while we used 25 lines in the case of HD 283572 as a
consequence of the lower quality of its spectrum.
\begin{figure}[!t]
\begin{center}
\scalebox{0.5}{
\includegraphics{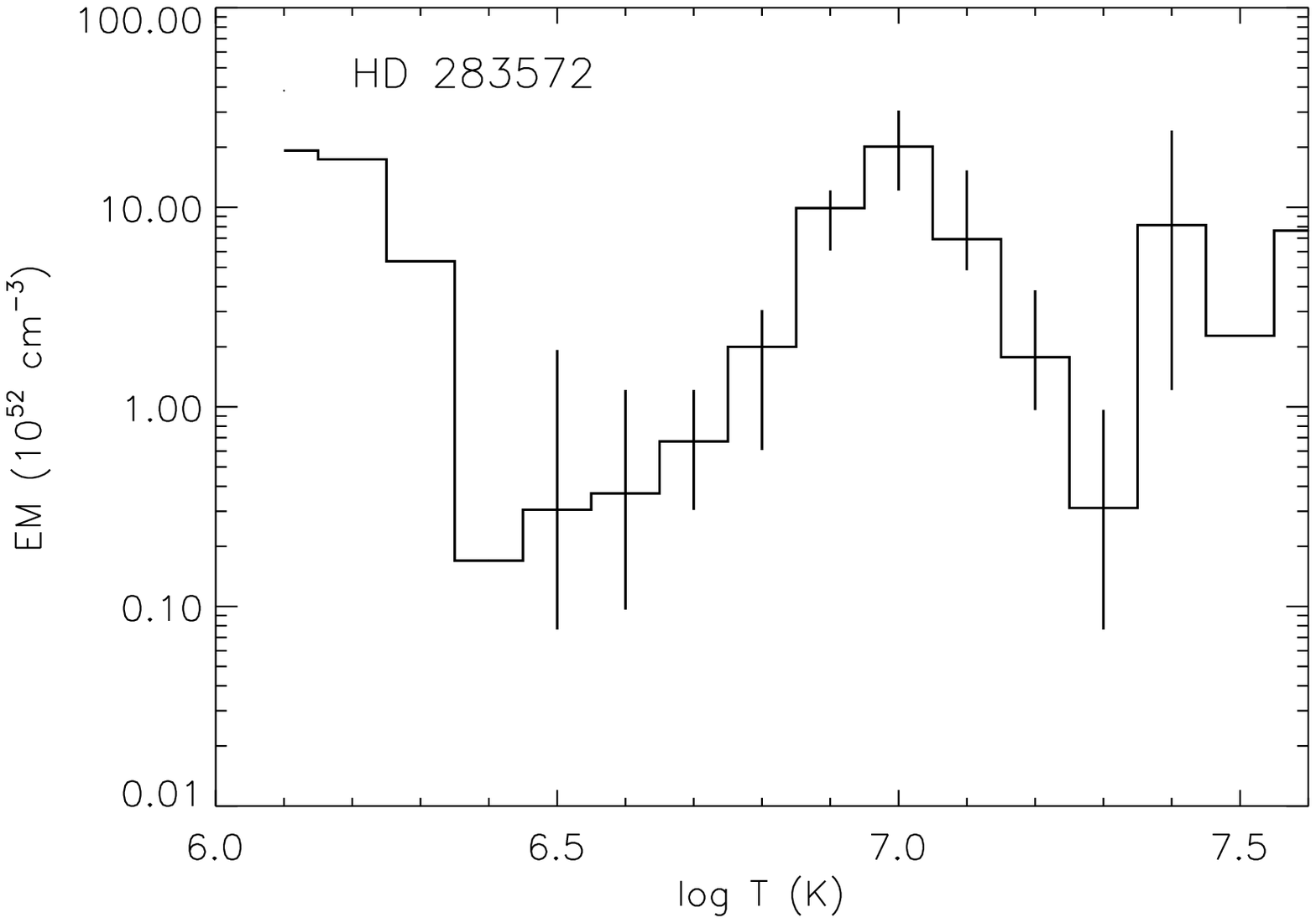}}
\scalebox{0.5}{
\includegraphics{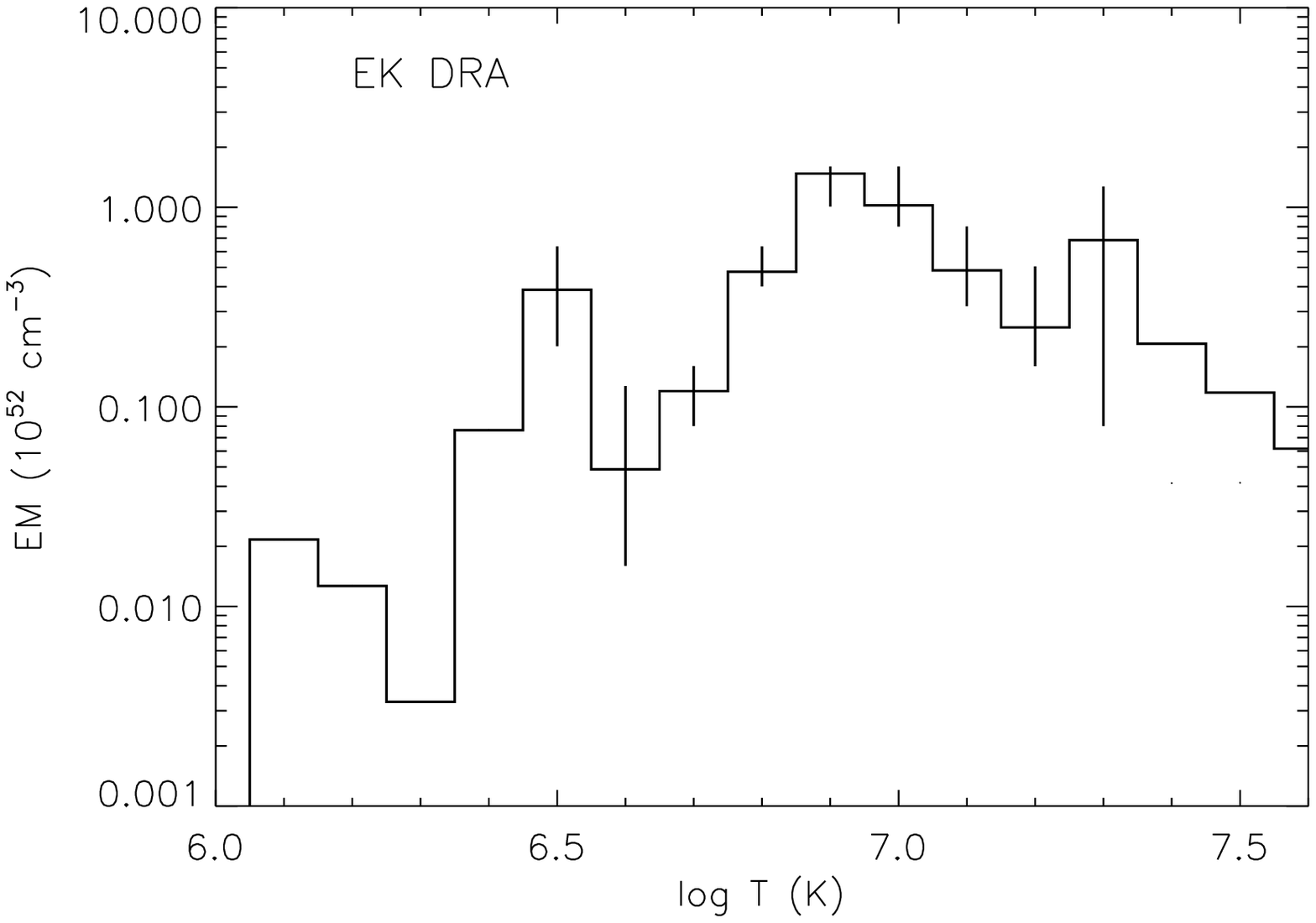}}
\scalebox{0.5}{
\includegraphics{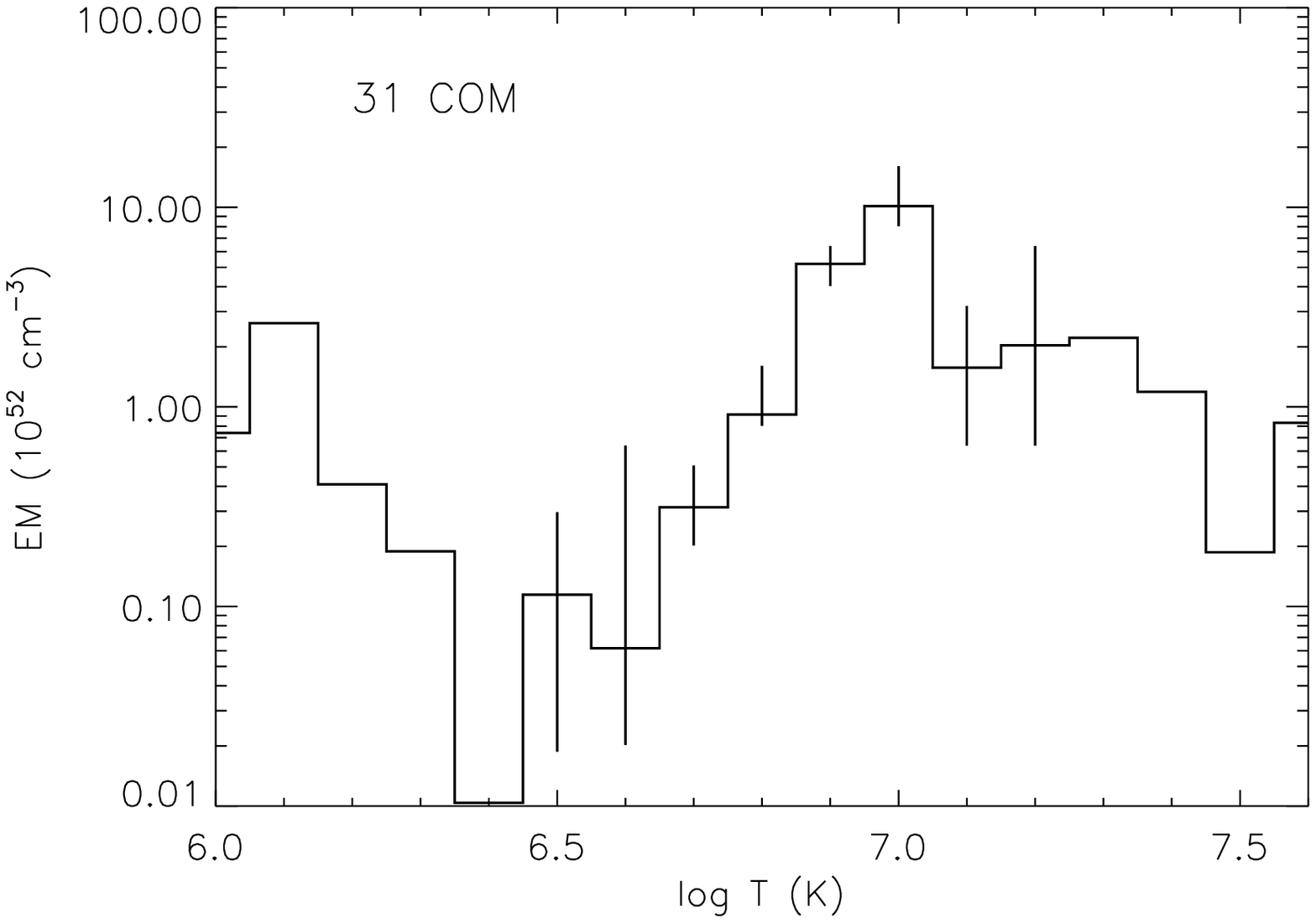}}
\caption{Distributions of emission measure derived from RGS data. Values without
error bars are not statistically constrained by the MCMC algorithm. Note the
different ordinate scale in the plot of EK Dra with respect to those of 
HD 283572 and 31 Com.} 
\label{fig:dem}
\end{center}
\end{figure}

\begin{figure}[t!]
\begin{center}
\scalebox{0.5}{
\includegraphics{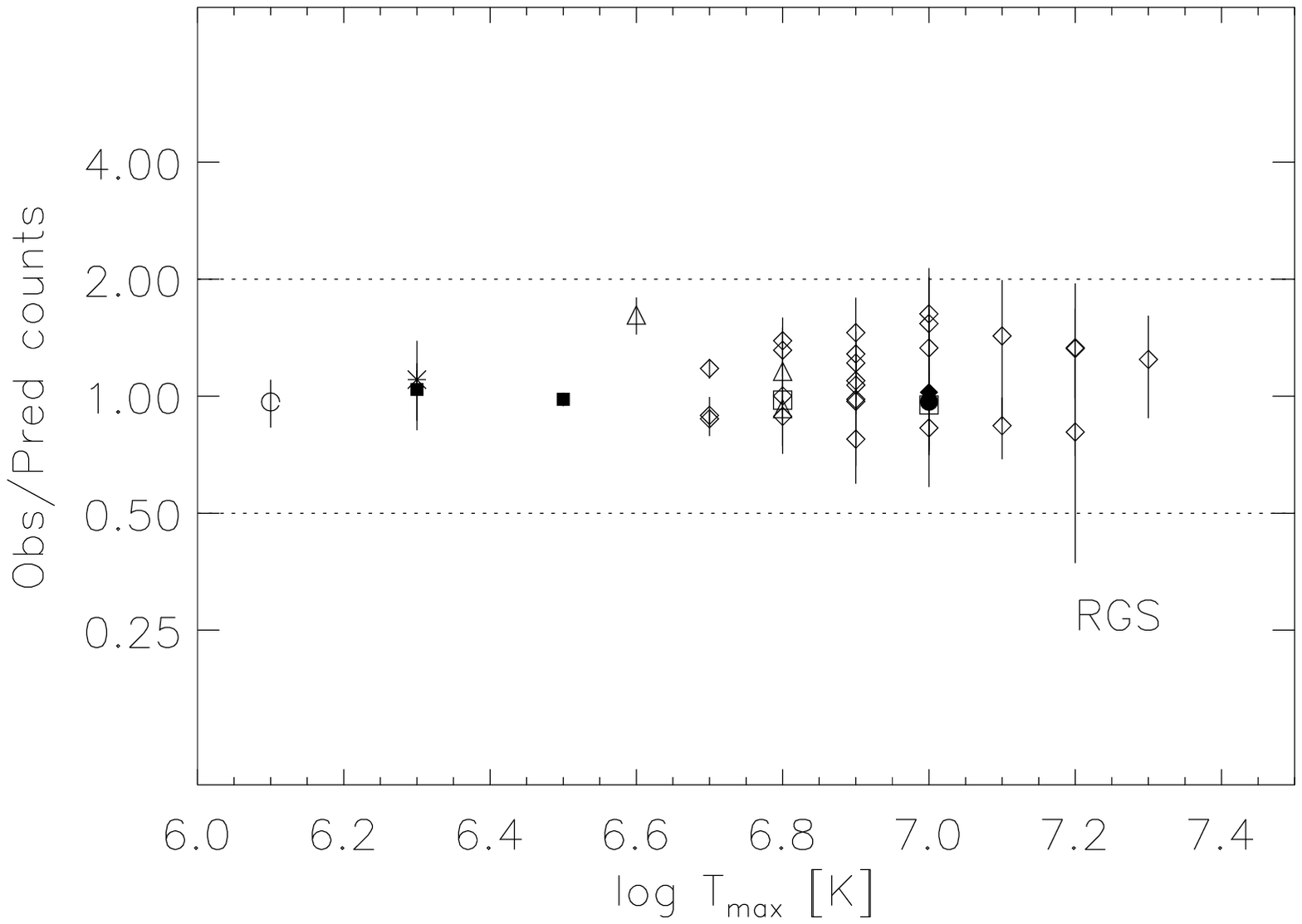}}
\caption{Comparison between observed fluxes and the fluxes predicted with the
$EMD$ model, for lines used in the EM reconstruction of EK Dra; Fe: open
diamonds, Ne: triangles, Mg: open squares, Si: filled diamond, Ni: filled
circle, O: filled squares, N: asterisk, C: open circle.}
\label{fig:check_flx}
\end{center}
\end{figure}

\begin{figure*}[p]
\begin{center}
\scalebox{0.5}{\includegraphics{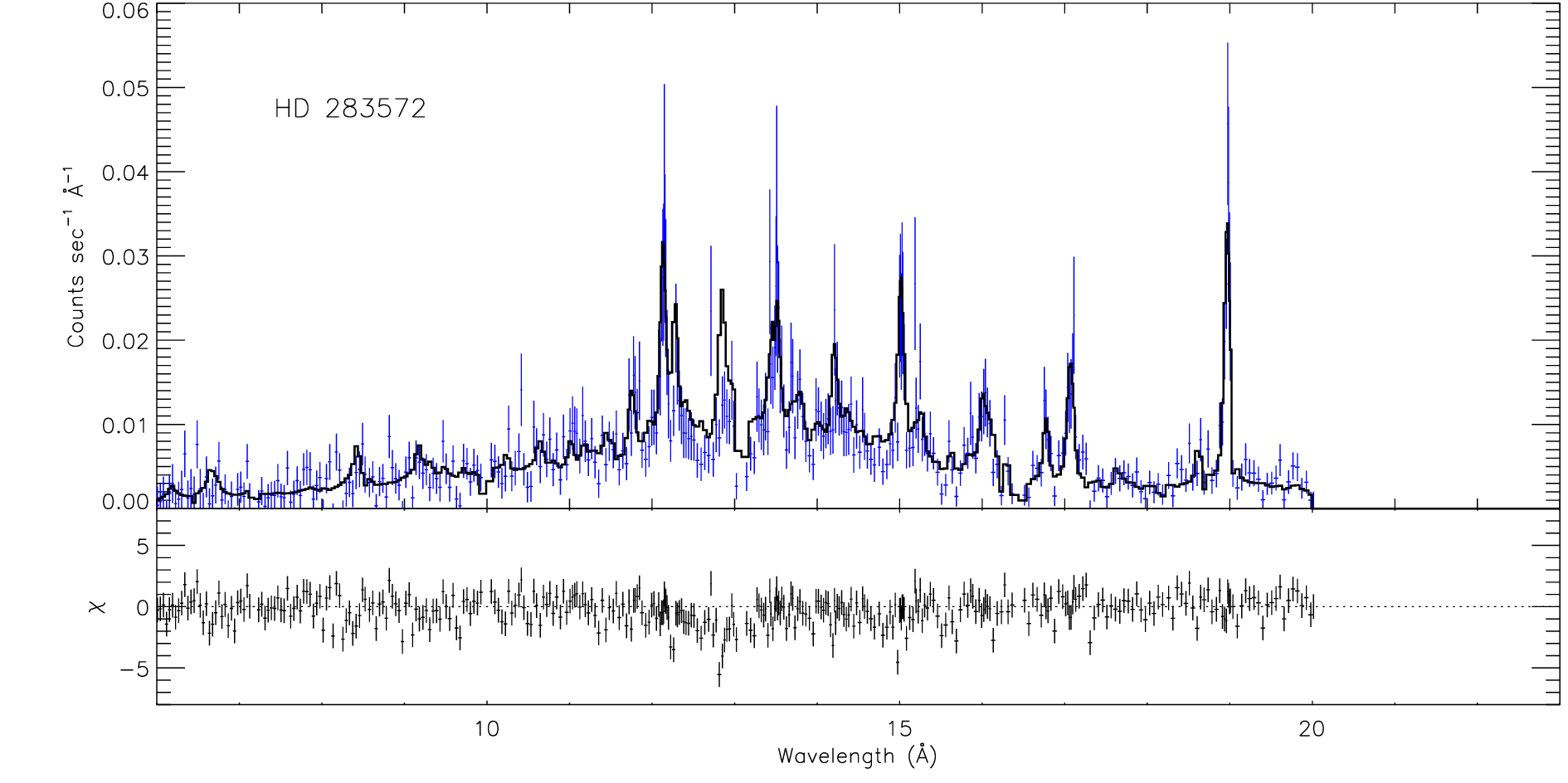}}
\scalebox{0.5}{\includegraphics{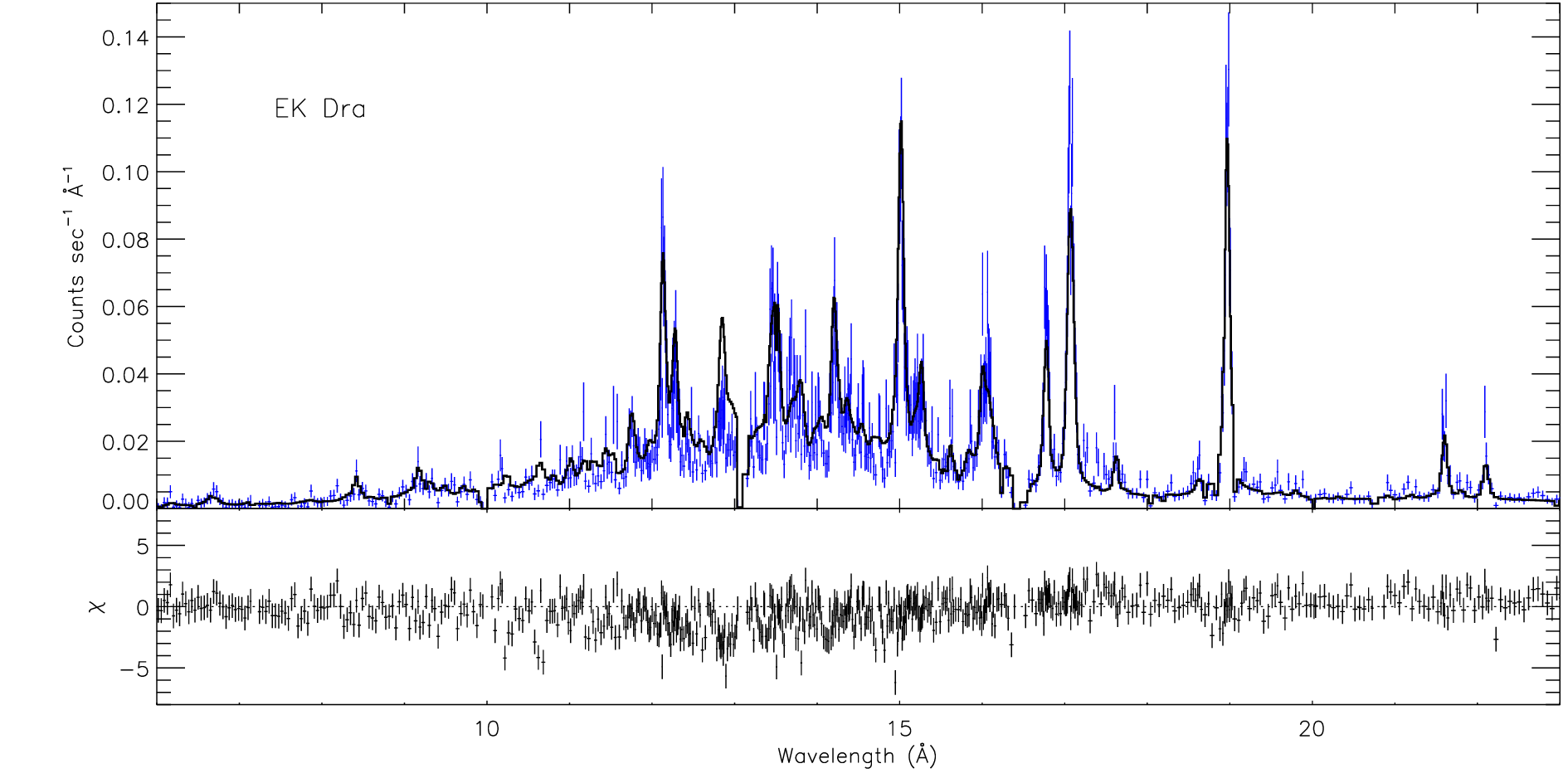}}
\scalebox{0.5}{\includegraphics{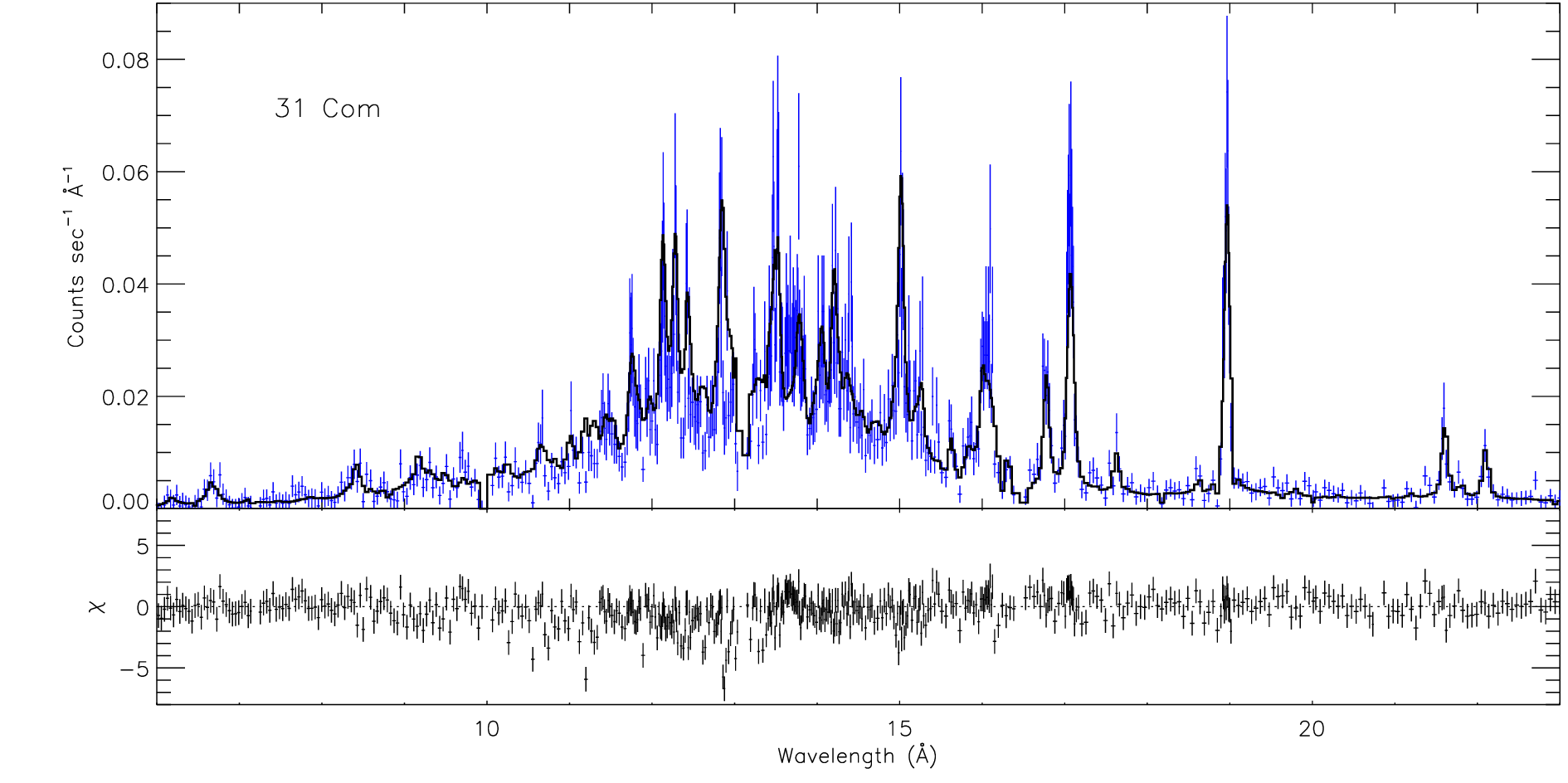}}
\caption{Model spectra compared to the original RGS spectra.}
\label{fig:model_spectra}
\end{center}
\end{figure*}

The derived $EMD$s are plotted in Fig. \ref{fig:dem}; note that the algorithm
we used is not able to constrain statistically the values of the emission
measure in all the temperature bins. We show in Fig. \ref{fig:check_flx} the
observed-to-predicted fluxes for the case of EK Dra, which is representative of
the spread of these ratios obtained in our analyses, and in Fig.
\ref{fig:model_spectra} we compare the observed spectra and the
model spectra generated with the solutions ($EMD$ and abundances) found in
this work. The elemental abundances are shown in Table \ref{tab:abund_RGS}; we
estimated the iron abundances (relative to the solar value) of HD 283572, EK Dra
and 31 Com at $0.7 \pm 0.2$, $1.2 \pm 0.2$ and $1.4 \pm 0.2$, respectively.
Table \ref{tab:pressure} reports the ratios\footnote{$r$, $i$ and $f$ denote the
fluxes of the resonance, intercombination and forbidden lines.} $R=f/i$ and
$G=(f+i)/r$ relative to the \ion{O}{vii} triplet, and the estimates of electron
temperatures, densities and pressures, averaged over the region where the
triplet forms, using the theoretical curves by \citet{SmithAPED2001}.

\begin{figure}[!h]
\begin{center}
\scalebox{0.5}{
\includegraphics{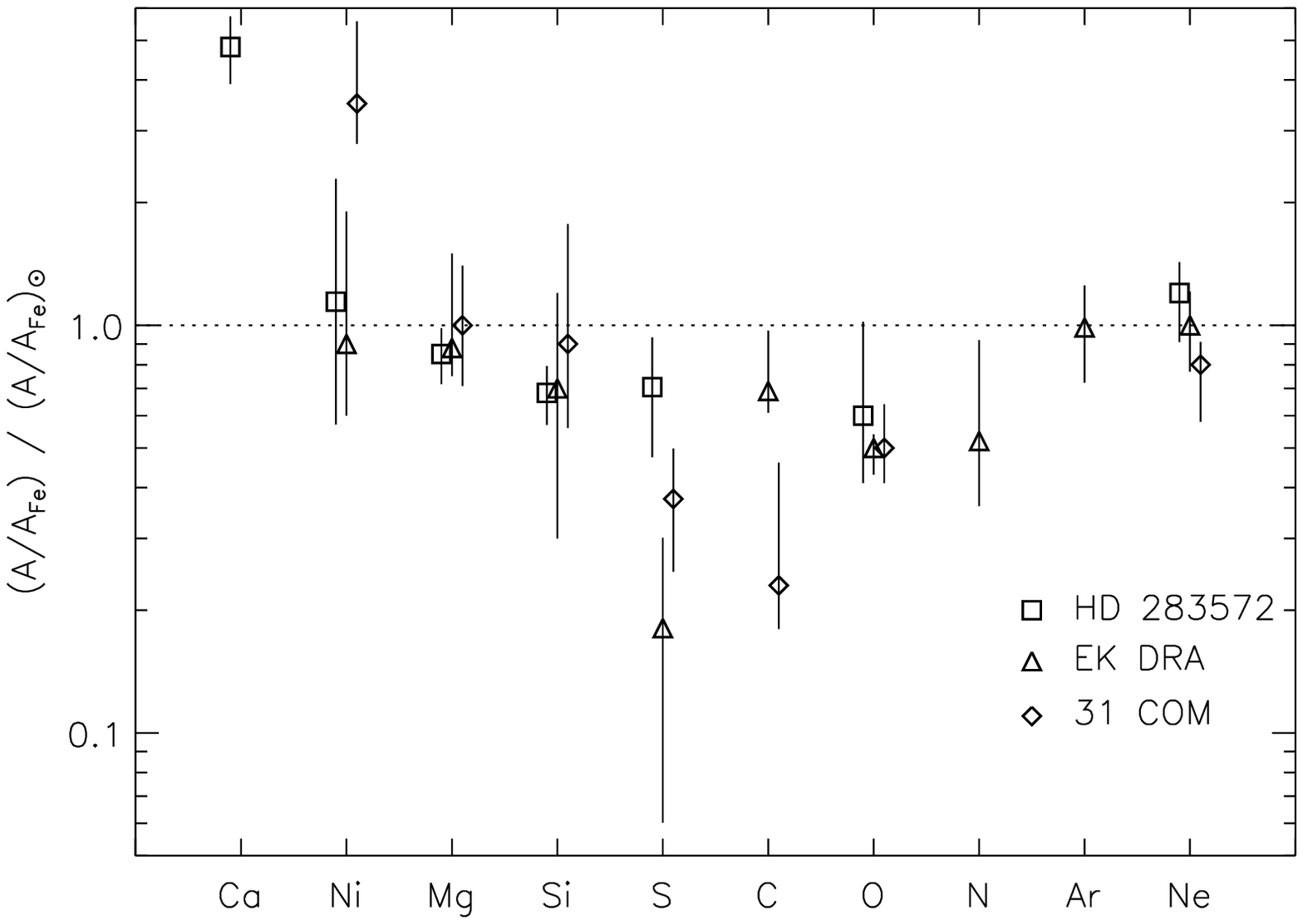}}
\caption{Element-to-iron abundance ratios, relative to the solar photospheric
values \citep{Grevesse1992}, for HD 283572 (squares), EK Dra (triangles) and 
31 Com (diamonds). The elements are ordered by increasing FIP.} 
\label{fig:FIP}
\end{center}
\end{figure}

In Fig. \ref{fig:FIP} we show the element-to-iron abundance ratios for the three
stars, ordering the elements for increasing First Ionization Potential (FIP).
Whenever both EPIC {\sc pn}- and RGS-based estimates were available, we always
report the latter in the plot, because we consider the values derived with the
RGS the most accurate. It is worth noting that, despite widely differing
methods employed to derive elemental abundances, we obtained consistency between
the {\sc pn}- and RGS-derived abundance ratios, except for Ne in the case of 31
Com (the value indicated by the {\sc pn} is two times larger than the RGS one)
and for Ni in the case of EK Dra and HD 283572 (the values obtained with the 
{\sc pn} are larger than the RGS ones by factors of 2 and 4 respectively).
\begin{center}
\begin{table*}[t!]
\caption{Ratios between elemental and iron coronal abundances, relative to
the solar photospheric ratios \citep{Grevesse1992}, derived from RGS data;
errors are at 68\% confidence level. For completeness, we also report 
the absolute iron abundance. The {\sc pn}-derived values are shown for 
purpose of comparison. For each star, the number of lines used for the 
$EMD$ reconstruction is reported (the number of lines of a given element
is shown in parenthesis near the relevant (RGS) abundance value).}
\vspace{0.3cm}
\begin{center}
\begin{tabular}{lcccccc} \hline\hline
             &   \multicolumn{2}{c}{HD 283572}       &   \multicolumn{2}{c}{EK Dra}        &   \multicolumn{2}{c}{31 Com}    \\ 
	     &              RGS   &  {\sc pn}   &      RGS   &  {\sc pn}   &  RGS   &   {\sc pn}   \\ \hline
   C/Fe         &                       &  &  $0.69_{-0.08}^{+0.28}$ (1)&  & $0.24_{-0.07}^{+0.22}$  (1) & \\ 
   N/Fe         &                       &  &  $0.52_{-0.16}^{+0.4}$ (1) &  &                             & \\ 
   O/Fe         &  $0.6_{-0.2}^{+0.4}$ (2) & $0.64\,\pm\,0.04$ &  $0.50_{-0.07}^{+0.04}$ (3) & $0.42\,\pm\,0.02$ &  $0.49_{-0.08}^{+0.15}$ (2) & $0.38\,\pm\,0.02$\\ 
   Ne/Fe        &  $1.2_{-0.3}^{+0.23}$ (2) & $1.24\,\pm\,0.09$ &  $1.00_{-0.23}^{+0.21}$ (3) & $1.00\,\pm\,0.05$ & $0.78_{-0.3}^{+0.13}$ (2)  & $1.53\,\pm\,0.09$\\ 
   Mg/Fe        &  & $0.86\,\pm\,0.14$  & $0.88_{-0.13}^{+0.6}$ (2) & $1.04\,\pm\,0.07$ & $1.0_{-0.3}^{+0.4}$ (2)  & $1.27\,\pm\,0.08$  \\ 
   Si/Fe        &  & $0.68\,\pm\,0.11$ &  $0.7_{-0.4}^{+0.5}$ (1) & $0.71\,\pm\,0.07$ & $0.9_{-0.3}^{+0.9}$ (1)   & $0.80\,\pm\,0.07$ \\ 
   Ni/Fe        &  $1.2_{-0.6}^{+1.0}$ (2) & $4.1\,\pm\,0.3$ & $0.9_{-0.3}^{+1.0}$ (1) & $2.17\,\pm\,0.24$ & $3.5_{-0.7}^{+2.1}$ (6)  & $2.7\,\pm\,0.2$  \\ \hline
   Fe           &  $0.7\,\pm\,0.2$ (19) & $0.37\,\pm\,0.01$ & $1.2\,\pm\,0.2$ (24) & $0.83\,\pm\,0.01$ & $1.4\,\pm\,0.2$ (27)  & $1.54\,\pm\,0.02$  \\ \hline
total lines  &            25           &  &           36         & &           41      &       \\ \hline
\end{tabular}
\end{center}
\label{tab:abund_RGS}
\end{table*}
\end{center}

The patterns of abundances vs. FIP are similar in the cases of 31 Com and 
HD 283572, with an initial decrease (with respect to solar photospheric values)
down to a minimum around carbon, followed by increasing abundances for elements
with higher FIP ($> 11$\,eV). This pattern is also similar to what was found
for the young active star AB Dor by \citet{Sanz2003}, but it is less evident in
the case of EK Dra. Note that 31 Com and EK Dra have iron abundances differing
from that of HD 283572 by about a factor of 2, hence the pattern of abundances
vs. FIP appears to be almost independent of the global coronal metallicity. 

\begin{center}
\begin{table}[b]
\caption{Pressure estimates with \ion{O}{vii}.}
\vspace{0.3cm}
\begin{center}
\scriptsize
\begin{tabular}{lccccc} \hline\hline
       &  $R$ &      $n_{e}$ (range)   & $G$ &       $T$       &     $P$ (range)     \\ 
       &      & ($10^{10}$\,cm$^{-3}$) &     &  ($10^{6}$\,K)  &  (dyn cm$^{-2}$)    \\ \hline
EK Dra & $3.0 \pm 1.7$ & $1$ ($< 7$)    & $0.93 \pm 0.25$ & $1.5^{+2.0}_{-0.5}$ & 4 ($<70$) \\
31 Com & $2.0 \pm 1.4$ & $3$ ($0.6-20$) & $0.94 \pm 0.36$ & $1.5^{+2.5}_{-0.7}$ & 13 ($1.5-220$) \\ \hline
\end{tabular}
\end{center}
\label{tab:pressure}
\end{table}
\end{center}

\section{Discussion}
\label{Discussion}

XMM-\emph{Newton} data allowed us to derive the plasma emission measure
distributions for our three targets and their coronal elemental abundances; in 
particular, the $EMD$ of HD 283572 has been derived here for the first time
using a high-resolution spectrum. Our results are sufficiently well determined
and homogeneous for the purpose of a detailed comparison of the coronal
properties of the selected stars. We recall that these stars are in different
evolutionary stages, but share the characteristic of being active (high X-ray
luminosity) G-type stars. Our analysis has confirmed that the three stars 
have very hot coronae, with similar average temperatures ($\sim 11-12$\,MK for
EK Dra and 31 Com, and $\sim 16$\,MK for HD 283572). 

A remarkable result of this work is the close similarity of the emission measure
distributions of HD 283572 and 31 Com, which have similar
$L_{\rm X}$ as well. Both distributions have a well-defined peak at 
$T_{\rm p}=10^7$\,K and, in the range $\log T \sim 6.5-7$, they are proportional
to $\sim T^5$, where the exponent of the power law has a formal confidence
interval between $\sim 3.4$ and $\sim 6.6$; there are also indications of a
significant amount of plasma at temperatures hotter than $T_{\rm p}$ (up to
$\log T \sim 7.6$) and, at least in the case of 31 Com, in the range 
$\log T \sim 6-6.2$. We recall that we are not able to statistically constrain
the emission measure in all the temperature bins, and hence to get information
on the exact shape of the distributions below $\log T \sim 6.5$ and above 
$\log T \sim 7.3$, yet the presence in both stars of a non-negligible amount of
plasma up to $\log T \sim 7.6$ has been verified, as described in Sect.
\ref{EMR}, through the correct prediction of the \ion{Fe}{xxiii-xxiv} line
fluxes and by comparison with the high-energy tail of the observed EPIC spectra
(App. \ref{check_EK_hot_tail}), while the correct prediction of the
\ion{O}{vii-viii} lines allowed us to verify the presence of cool plasma down to
$\log T \sim 6$ in the  $EMD$ of 31 Com (the \ion{O}{vii} line is not available
in the spectrum of HD 283572). Note, also, that the shape of the constrained
part of the distribution of 31 Com and the presence of significant emission
measure at $\log T \sim 6-6.2$ suggest that a minimum in the $EMD$ of this star
occurs around $\log T \sim 6.4-6.5$, while some caution is needed for the case
of HD 283572.

The $EMD$ of EK Dra is, on average, about one order of magnitude lower
than the two previous ones. The distribution has a maximum at $\log T=6.9$,
with a more gradual decrease towards higher $T$ than in the two previous cases.
Using ASCA/EUVE data, \citet{Guedel_EK1997} derived for EK Dra an $EMD$
essentially bimodal, with two significant peaks near 7\,MK and 18\,MK. While we
also find little plasma at temperatures below $\sim 3$\,MK and our value of
$T_{\rm p}$ ($\sim 8$\,MK) is roughly consistent with their first peak, we do
not find either a deep minimum around 10\,MK or strong evidence for a second
maximum at 18\,MK.

\begin{figure}[t]
\begin{center}
\scalebox{0.5}{
\includegraphics{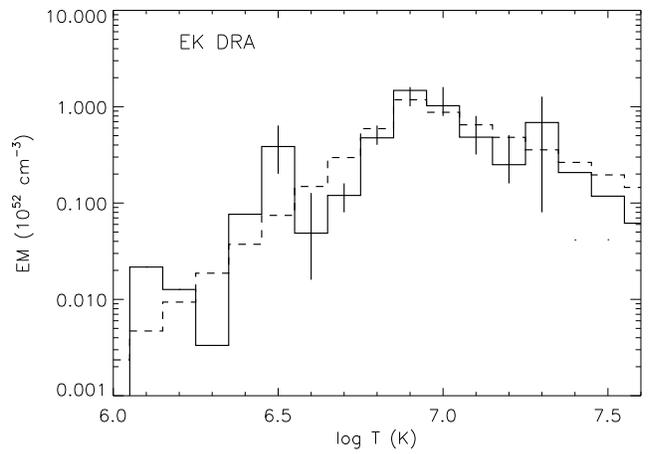}}
\caption{Emission measure distribution of EK Dra reconstructed in this work
using RGS data and the MCMC algorithm (solid line) and the smoothed solution
whose shape is proportional to $T^3$ for $T<10^7$\,K and to $T^{-1.3}$ for
$T>10^7$\,K (dashed line).} 
\label{fig:EK_T5vsT3}
\end{center}
\end{figure}

\begin{figure}[t]
\begin{center}
\scalebox{0.5}{
\includegraphics{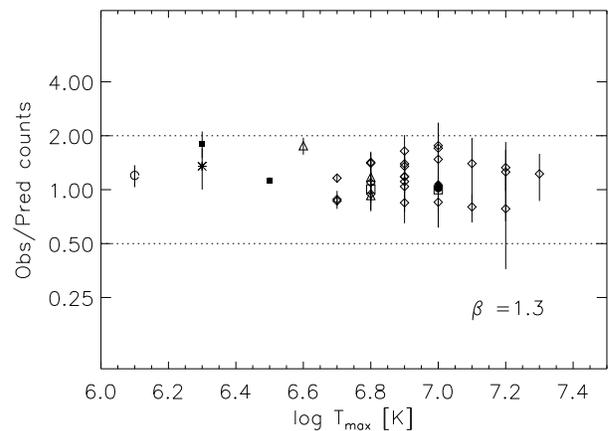}}
\caption{Comparison between observed fluxes and the fluxes predicted with the
smoothed $EMD$ (dashed line in Fig. \ref{fig:EK_T5vsT3}), for lines used in
the EMD analysis of EK Dra; Fe: open diamonds, Ne: triangles, Mg: open squares,
Si: filled diamond, Ni: filled circle, O: filled squares, N: asterisk, C: open
circle.} 
\label{fig:check_EK_T3_1.3}
\end{center}
\end{figure}

It is not straightforward to derive a low-$T$ slope for this distribution,
essentially due to the secondary peak at $\log T=6.5$. If we exclude this
temperature bin, on the grounds that it belongs to a cooler population of
coronal structures (see below), a fitting in the range $\log T=6.6-6.9$,
yields a slope of $5.3 \pm 1.7$: this is compatible, within errors, with the
slopes derived for HD 283572 and 31 Com. Instead, if we consider the secondary
peak as a fluctuation produced by the emission measure analysis, the slope 
in the range $\log T=6.5-6.9$ turns out to be $3.0 \pm 1.2$. To address the
issue of the $EMD$ slope, we have investigated if a solution smoother than the
one presented in Fig. \ref{fig:dem} can give a good description of the observed
line fluxes as well. Assuming a distribution with its ascending part 
($\log T<6.9$) proportional to $T^3$ and with the high-temperature tail 
($\log T>6.9$) decreasing as $T^{-\beta}$, we searched for the minimum $\chi^2$
on the subset of the iron lines, by varing the exponent $\beta$ and a global
renormalization factor. The minimum $\chi^2$ is obtained for $\beta = 1.3$ (the
relevant $EMD$ is shown in Fig. \ref{fig:EK_T5vsT3} together with the
reconstructed one), and the observed fluxes for all the selected lines are quite
acceptably reproduced (Fig. \ref{fig:check_EK_T3_1.3}, to be compared with 
Fig. \ref{fig:check_flx}). Note also that we used, for this test, the relative
element abundances reported in Table \ref{tab:abund_RGS}. However this solution
over-predicts the \ion{O}{viii} 18.97\,\AA/\ion{O}{vii} 21.60\,\AA\ line ratio:
while the observed ratio of line counts is $R_{\rm O, obs} = 10.0\pm 1.7$ 
(1 $\sigma$ error), the ratio predicted by the smooth solution is 
$R_{\rm O} \sim 16$, against $R_{\rm O} = 10.6$ predicted by the MCMC solution.
This line ratio does not depend on the oxygen abundance, but it is especially
sensitive to the shape of the distribution at low temperatures ($\log T<6.8$),
hence its predicted value could be lowered by a small enhancement of the
emission measure at $\log T=6.0-6.3$ without affecting significantly the fluxes
of the iron lines\footnote{In this case an adjustment of the C and N abundances
would be required.}. Interestingly, our smooth solution is similar to that found
recently by \citet{Telleschi2004} using the same XMM data, but a different
inversion method. Thus, we are not able to get strong constraints on
the low-$T$ slope of the $EMD$ of EK Dra and further investigation is needed.

We want to interpret the shape of the bulk of the emission measure
distributions, around $10^7$\,K, in terms of loop structures. First, we note
that, at this high coronal temperature, the pressure scale height $H_{\rm p}$ in
each of the three stars is of the order of the corresponding radius (Table
\ref{tab:H_p}). We assume that the structures responsible for the observed
emission of each star have characteristic sizes smaller than the relevant
pressure scale height. In fact, for 31 Com we showed in \citet{Scelsi2004} that
loop heigths larger than the stellar radius would be hardly compatible with the
absence of variability in the emission of this star, because of the very low
filling factor implied by this solution; in Appendix \ref{results_flare} we
analyze the flare on EK Dra and conclude that the height of the flaring loop,
which likely belongs to the family of structures dominating the X-ray emission,
is a few $10^{10}$\,cm; finally, \citet{Favata2001} analyzed a flare on HD
283572 and showed that the size of the involved structure is $\sim 0.3\,R_*$.
Under this hypothesis, the pressure is approximately uniform inside each loop,
implying that the emission measure distribution of a single loop depends only on
its maximum temperature $T_{\rm max}$ \citep{Maggio1996}, with a functional form
$EM(T) \propto T^{\alpha}$ for $T<T_{\rm max}$, with $\alpha=3/2$ in the case of
loops with constant cross-section and uniform heating. Considering that the
$EMD$ of the whole stellar corona is the sum of the $EM(T)$ of individual loops,
the total $EMD$ would be proportional to $T^{\alpha}$ for 
$T<{\rm min}\{T_{\rm max}\}$; hence, following the approach by
\citet{Peres2001}, we interpret the constrained part of the $EMD$s of
HD 283572 and 31 Com as due to a population of loops, each of them having 
$EM(T) \propto T^5$, since the $EMD$s of these stars are approximatively
power-laws (with exponent $\sim 5$) in the temperature range mentioned above.
Consequently, the simplest interpretation we derive from the comparison of the
emission measure distributions of HD 283572 and 31 Com (Fig. \ref{fig:dem}, see also Fig. \ref{fig:all_dem} below) is that the coronae of
these stars are very similar in terms of dominant coronal structures, in spite
of their different evolutionary phases and gravities, as well as coronal
abundances. Since this latter parameter plays an important role in the energy
balance through the radiative losses, we might expect that different abundances 
result in different temperature and density profiles along a loop, and hence in
different coronal $EMD$s; however, this is not the case in these two stars.
Moreover, we stress that HD 283572 and 31 Com show similar $EMD$s in spite of
the difference in X-ray surface flux by about one order of magnitude. In
conclusion, we infer that all the above parameters have only a minor role in
determining the properties of the $EMD$ of these stars, which instead appear to
be mainly determined by the high and nearly identical X-ray luminosity.

The high index ($\sim 5$) of the power law which best approximates the ascending
part of the $EMD$s of these stars also suggests that the dominant coronal
loops of very bright sources (with $L_{\rm X}\sim 10^{31}$\,erg\,s$^{-1}$) may
have different properties to the solar ones. In such stars, the physical
processes that lead to emission measure distributions significantly steeper than
those observed in low-luminosity stars, such as the Sun or $\alpha$ Cen
\citep{Drake1997}, still remain to be understood; as discussed in
\citet{Scelsi2004}, a possible interpretation of such steep slopes, which should
also characterize the $EM(T)$ of the single structures (see above), is that the
heating of the coronal loops is located mainly at their footpoints 
\citep{Testa2003,Testa2004}. Another possibility are expanding loops,
although quite extreme expansion factors are needed \citep{Schrijver1989}; also,
both effects might be at work.

The interpretation of the $EMD$ of EK Dra is more uncertain. If we assume that
the distribution reconstructed in this work and shown in Fig. \ref{fig:dem} is a
good approximation of the actual $EMD$ of this star, then a family of hot loops
with $EM(T) \propto T^5$ appears to be responsible for the bulk of the
distribution around $10^7$\,K, while the emission measure at $\log T < 6.6$ may be due to a cooler family of loops whose properties we are not able to
investigate with the available data, because of the limited information on the
low-temperature plasma; otherwise, if a smoother distribution applies 
($EMD \propto T^3$, Fig. \ref{fig:EK_T5vsT3}), the structures dominating
this corona might have a less steep profile of the emission measure vs.
temperature with respect to the cases of HD 283572 and 31 Com, yet steeper than
the $T^{3/2}$ slope characterizing uniformly-heated loops with constant
cross-section.
\begin{center}
\begin{table}[t!]
\caption{Stellar radii and pressure scale heigths calculated for 
$T\sim 10^7$\,K.}
\vspace{0.3cm}
\begin{center}
\begin{tabular}{lcc} \hline\hline
           &          $R$           &  $H_{\rm p}$  \\ 
           &          (cm)          &          (cm)             \\ \hline
HD 283572  & $\sim 3\times 10^{11}$ &   $\sim 3\times 10^{11}$  \\
EK Dra     &   $6.6\times 10^{10}$  &     $4\times 10^{10}$     \\
31 Com     &   $6.5\times 10^{11}$  &        $10^{12}$          \\ \hline
\end{tabular}
\end{center}
\label{tab:H_p}
\end{table}
\end{center}

In Fig. \ref{fig:all_dem}, the $EMD$s of the three studied stars are shown
together with the emission measure distribution of the Sun \citep{Peres2000}
and \object{$\xi$ Bootis} \citep{Laming1999}, the latter being a G-type star of
intermediate activity, with $L_{\rm X}\sim 10^{29}$\,erg\,s$^{-1}$. While the $EMD$ of
the Sun peaks at $T_{\rm p} \sim 10^{6.2}$\,K, with an ascending part 
($\log T \sim 5.7-6.2$) proportional to $T^{3/2}$, and shows no significant
amount of plasma at temperatures above $\sim 10^{6.7}$\,K, the $EMD$ of $\xi$
Bootis is intermediate between those of the Sun and our active stars, in terms
both of $T_{\rm p}$ and of the overall amount of emitting plasma, as well as
with regard to the steepness of the $EMD$ preceding its peak. Actually, $\xi$
Bootis is a double star (G8+K4V), but the X-ray emission is thought to be
dominated by the primary G8 \citep{Schmitt1997}, and we have included it in this
picture because it is one of the very few stars of intermediate activity whose
$EMD$ has been reconstructed from high-resolution spectra.
\begin{figure*}[!t]
\begin{center}
\scalebox{0.8}{
\includegraphics{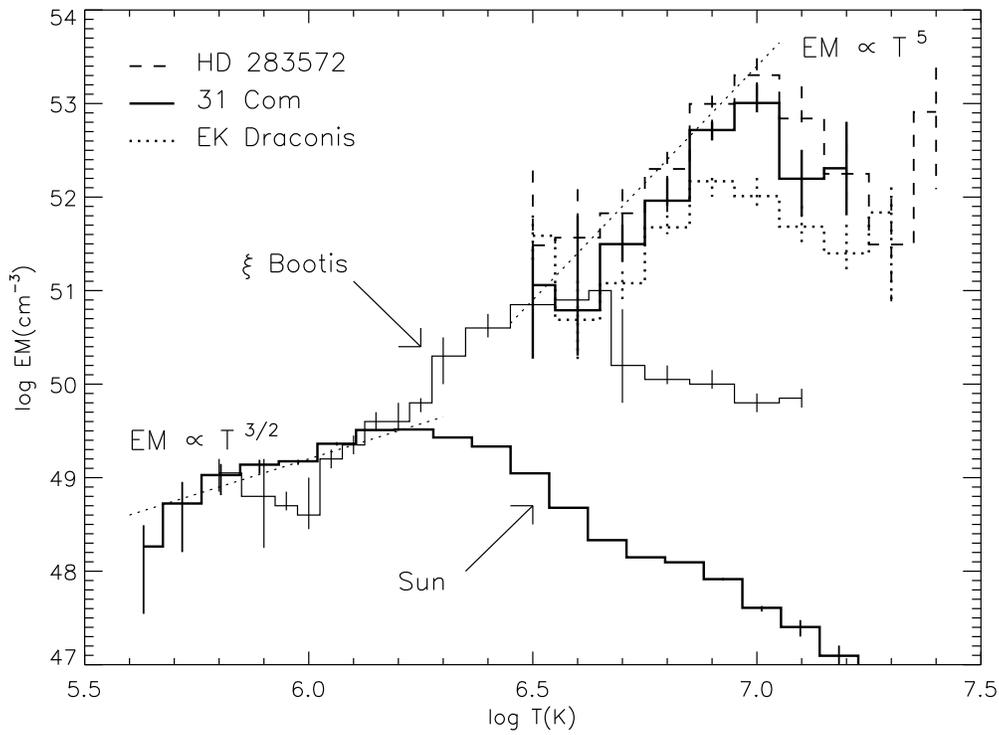}}
\caption{Emission measure distributions of the coronal plasma of the quiet Sun 
\citep[][from $Yohkoh$/SXT data]{Peres2000}, $\xi$ Bootis
\citep[][from ASCA/EUVE data]{Laming1999}, EK Dra, 31 Com and HD 283572 (this
work). For each of the three active stars, only the constrained part of the
$EMD$ is shown.} 
\label{fig:all_dem}
\end{center}
\end{figure*}

The comparison shown in Fig. \ref{fig:all_dem} between the $EMD$s of stars with
increasing luminosity, going from the quiet Sun to HD 283572, suggests a
transition in the steepness of the distribution, which, in turn, may reflect
changes of the properties of the dominant coronal loops. Fig. \ref{fig:all_dem}
is in agreement with the hypothesis of increasing steepness with increasing
$L_{\rm X}$ reported in \citet{Bowyer2000}; on the other hand, their picture is
not reflected completely by our results which do not indicate, at coronal
temperatures, a monotonic increase of the $EMD$ up to its peak, at least in
the cases of the bright star 31 Com and possibly also of HD 283572; in fact, the
XMM data available for these two stars suggest the presence of plasma at 
$T\sim 10^6$\,K and a minimum of the $EMD$ around $T\sim 10^{6.5}$\,K. 

Finally, it is important to note that the sample is still limited, and new
observations are required to have a more complete scenario, as well as to study
in greater detail the ''cool'' and ''hot'' tails of the emission measure
distributions of very active stars. In this respect, the recent work by
\citet{Telleschi2004} provides complementary results which may help to bridge 
the gap between solar-type stars and stars with very high activity levels.

\appendix

\section{Checking the high temperature tail of the $EMD$s}
\label{check_EK_hot_tail}

\begin{figure}[t]
\begin{center}
\scalebox{0.5}{
\includegraphics{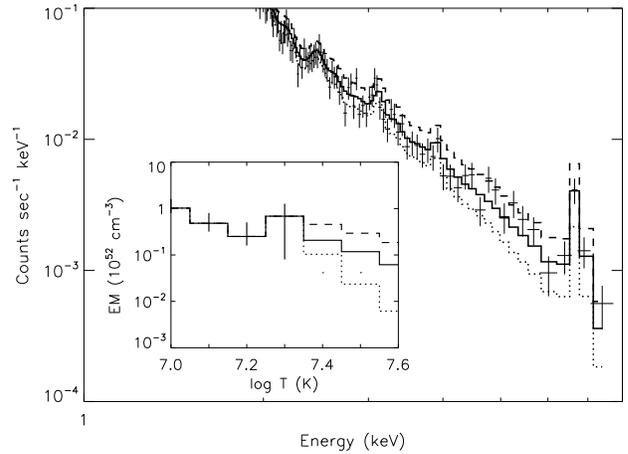}}
\caption{EPIC {\sc pn} spectrum of EK Dra (crosses) and the model spectrum
derived from the reconstructed $EMD$ (solid line); the dashed line is the model
spectrum obtained in the case of a distribution equal to the reconstructed one
up to $\log T =7.3$, but with a flatter hot tail drawn with the dashed line in
the inset; the model spectrum shown with the dotted line, instead, is relevant
to a steeper decrease of the hot tail.}  
\label{fig:EK_hot_tail}
\end{center}
\end{figure}

As stated at the end of Sect. \ref{EMR}, we checked the reliability of the hot
tails of the $EMD$s, not constrained by the MCMC method, by comparing the model
spectrum with the high-energy ($E > 2$\,keV) {\sc pn} spectrum, which is more
sensitive to very hot plasma (it contains, in particular, the complex around
6.7\,keV largely dominated by \ion{Fe}{xxv} and several spectral regions
dominated by the bremsstrahlung continuum). This comparison is shown in Fig.
\ref{fig:EK_hot_tail} for the case of EK Dra, similar results have been obtained
for HD 283572 and 31 Com. In this plot, the solid line is the {\sc pn} model
spectrum derived from the reconstructed $EMD$, whose hottest part is reported
(with the solid line) in the inset. In the high-energy tail of the spectrum
($2-8$\,keV) the model fits well the data ($\chi^2_{\nu} \sim 0.98$, 88
$d.o.f.$), indicating reliability of the presence of a sizeable amount of plasma
in the high-temperature tail of the $EMD$. We have also examined both the cases
of a flatter hot tail with a total emission measure twice as large as the
previous case (approximatively proportional to $T^{-3/2}$, shown by the dashed
line in the inset), and of a tail which decreases as $T^{-6}$ (dotted line),
with a total emission measure four-fold lower than the first case. The
corresponding model spectra give a poor fit to the data ($\chi^2_{\nu} \sim 4$
and 1.4, respectively, consider that $P(\chi^2_{\nu} > 1.4) \sim 0.8\%$).

\section{Analysis of the flare on EK Dra}
\label{results_flare}

The flare observed on EK Dra (Fig. \ref{fig:lc}) has a duration of $\sim 10$\,ks
and is characterized by a rather wide ($\sim 3$\,ks) peak at $\sim 2$\,cts/s, 
subtracting the mean count rate of the quiescent phase, and a secondary maximum
at $\sim 0.5$\,cts/s which follows the initial decay phase. For the analysis of
this flare, we performed time-resolved spectroscopy of the EPIC {\sc pn} data
and employed the approach by \citet{Reale1997} to derive the size
of the flaring loop, assuming that the impulsively heated plasma was confined in
a single structure. We refer to that paper for a detailed explanation of this
method, and to \citet{RealeProxCen2004} for its application to a flare (on
\object{Proxima Centauri}) observed by XMM-\emph{Newton}.
\begin{center}
\begin{table*}[!t]
\caption{Best-fit values of temperature, emission measure and metallicity (with
90\% confidence errors) relevant to the thermal component of the model
describing the flaring plasma; for each segment, the (approximated) number of 
counts in the {\sc pn} spectrum is also reported.}
\vspace{0.3cm}
\begin{center}
\begin{tabular}{cccccccc} \hline\hline
        &        & \multicolumn{3}{c}{variable $z$} & & \multicolumn{2}{c}{fixed $z$} \\
\cline{3-5} \cline{7-8}
Segment & counts &    $T_{\rm obs}$      &          $EM$         &     $z$      & &           $T_{\rm obs}$      &          $EM$  \\ 
        &        &    ($10^{7}$\,K)      & ($10^{52}$\,cm$^{-3}$)&              & &          ($10^{7}$\,K)       & ($10^{52}$\,cm$^{-3}$) \\ \hline
r       &  3600  & $4.2^{+1.6}_{-1.0}$   &  $4.6^{+1.6}_{-1.5}$  &  $1 (<2.9)$  & &          $4.2^{+1.3}_{-0.9}$   &  $4.9^{+0.3}_{-0.6}$  \\
p1      &  4800  & $3.3^{+0.7}_{-0.6}$   &  $9.3^{+1.9}_{-1.8}$  &  $0.5^{+0.7}_{-0.4}$ & &  $3.4^{+0.5}_{-0.4}$   &  $8.5^{+0.6}_{-0.5}$  \\
p2$^{\,{\bf a}}$      &  4700  & $1.86^{+0.4}_{-0.25}$ &  $10.0^{+1.9}_{-1.8}$ &  $0.30^{+0.26}_{-0.17}$ & & $2.2^{+0.25}_{-0.3}$  &  $7.4^{+0.6}_{-0.5}$  \\
d1      &  4000  & $1.43^{+1.6}_{-0.23}$ &  $6.5^{+1.9}_{-1.7}$  &  $0.27^{+0.27}_{-0.15}$ & & $1.68^{+0.20}_{-0.25}$   &  $4.2^{+0.4}_{-0.7}$  \\
d2      &  4100  & $1.8^{+0.5}_{-0.3}$   &  $3.1 \pm 0.5$        &  $0.27^{\,{\bf b}}$     & & $2.0^{+0.6}_{-0.3}$   &  $2.1^{+0.4}_{-0.4}$     \\
p3      &  4800  & $2.4^{+1.4}_{-0.8}$   &  $2.4^{+1.1}_{-1.3}$  &  $0.48 (<2.8)$          & & $2.5^{+1.1}_{-0.6}$   &  $2.0^{+0.3}_{-0.3}$  \\ \hline
\multicolumn{8}{l}{$^{{\bf a}}$ The agreement between data and model in the case of fixed $z$ is not good ($P(\chi > \chi_{\rm obs}) = 0.6\%$).} \\
\multicolumn{8}{l}{$^{{\bf b}}$ Fixed to the value of segment d1.} \\
\end{tabular}
\end{center}
\label{tab:fit_segments}
\end{table*}
\end{center}
\begin{figure}[!h]
\begin{center}
\scalebox{0.5}{
\includegraphics{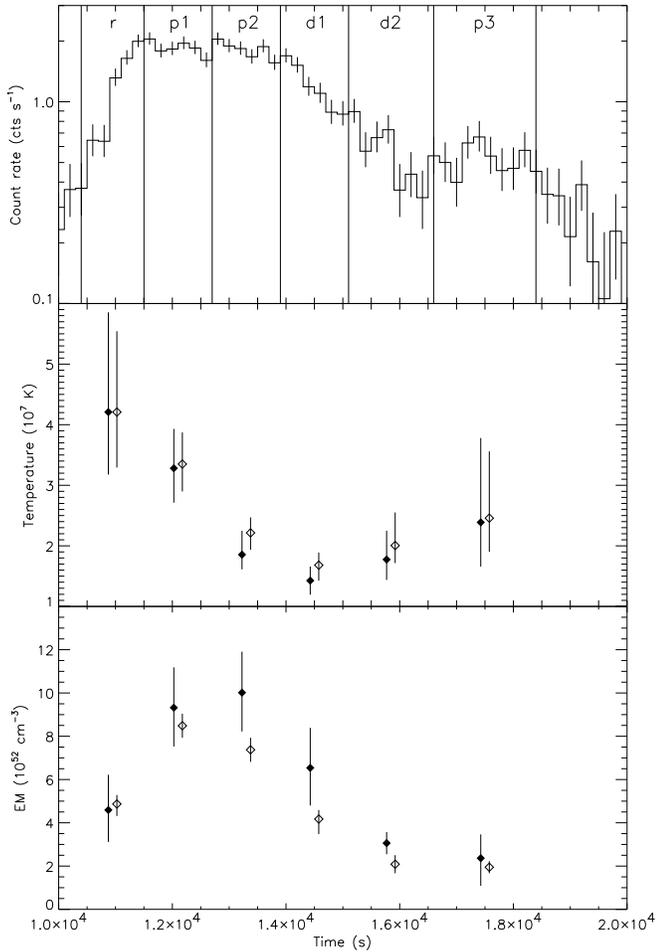}}
\caption{The upper panel shows the light curve of the flare on EK Dra, in the 
$0.3-10$\,keV band of the EPIC {\sc pn}, obtained by subtracting the average
quiescent emission from the total light curve in Fig. \ref{fig:lc}. The vertical
lines mark the phases of the flare. The middle and the lower panels show the
time evolution of the temperature and the emission measure, respectively. Filled
symbols are relevant to the model with variable global metallicity, open symbols
to the model with $z = 0.83$ solar.} 
\label{fig:segments}
\end{center}
\end{figure}
We divided the observation during the flare into 6 segments (Fig. 
\ref{fig:segments}), so as to have $\sim 4000-5000$ counts in the {\sc pn}
spectrum of each of them. These spectra were fitted in XSPEC using, as a
model, the fixed 3-T model in Table \ref{tab:fit_EPIC}, describing the quiescent
emission, plus a fourth component, which gives the temperature and the emission
measure of the flaring plasma. We made the fittings both using a variable global
metallicity ($z$) and fixing it to the quiescent value ($z = 0.83$ solar). The
results of the fittings are reported in Table \ref{tab:fit_segments} and the
evolutions with time of the temperature and the emission measure are shown in
the middle and lower panels of Fig. \ref{fig:segments}.

Unfortunately, the results of our analysis for this specific event are affected
by several uncertainties, hence the loop size we obtain is to be taken with
caution. 

The half-length $L$ of the loop is a function of the $e-$folding time of the
light curve, $\tau_{\rm LC}$, of the maximum temperature of the flaring plasma,
$T_{\rm obs, max}$, and of the slope $\zeta$ of the trajectory (during the
decay) in the density-temperature diagram. To derive $\tau_{\rm LC}$ we need the
light curve of the flare alone, which is the total light curve minus the
quiescent emission; yet, in our case, the latter component is comparable to the
flaring one and it is not constant, as already shown in Sect.
\ref{light_curves}. We approximated the quiescent emission with its
constant mean value in order to obtain the light curve of the flare, shown in
the upper panel of Fig. \ref{fig:segments}. This curve is not characterized by a
well-defined peak followed by an exponential decay, therefore it is not possible
to derive an accurate $e-$folding time from it. We estimated 
$\tau_{\rm LC}$ by evaluating the time, after the peak, when the count-rate has
fallen down to $1/e$ times the maximum value and we obtained 
$\tau_{\rm LC} \sim 4000$\,s.

Another difficulty arises from the second maximum\footnote{A second maximum was
observed in the light curve of several flares 
\citep[e.g.][]{Poletto1988,Pallavicini1990}. More recently,
\citet{RealeProxCen2004} modelled a very strong flare detected on Prox Cen and
showed that a second loop system, probably an arcade, is required to explain the
observed secondary maximum.} of the flare, which ''breaks'' the decaying phase. 
Since the statistic is not very high for such a kind of time-resolved spectral
analysis, we were able to derive the trajectory in the $n_{\rm e}-T$ diagram
during the decay and its slope only from two points (segments $d1$ and $d2$). 
The slope $\zeta$ is estimated to be $\sim 1.2$ using the results of the 
fittings with variable $z$, and $\sim 1$ in the second case; these values
indicate that sustained heating was present during the decay of the 
flare\footnote{If no heating is present during the decay $\zeta$ is $\sim 2$}.

The maximum observed temperature was evaluated to be 
$T_{\rm obs} \sim 4.2\times 10^7$\,K in both cases (segment $r$ in Table 
\ref{tab:fit_segments}); we conclude that the size of the flaring loop is of the
order of a few $10^{10}$\,cm, i.e. substantially smaller than the stellar
radius.

Finally, we derived the average temperature $T_{\rm eq}$ of the flaring
plasma at equilibrium, i.e. after the flare has totally decayed and the loop 
returned to its quiescent conditions, by fitting the temperature values of
segments $r$, $p1$, $p2$ and $d1$ with the function 
$T=A\,{\rm e}^{-t/\tau}\,+T_{\rm eq}$. Although the errors are rather large, we
obtained a minimum of $\chi^2$ for $T_{\rm eq}=6\times 10^6$\,K, in the case of
variable $z$, and $T_{\rm eq}=5\times 10^6$\,K, in the case of fixed $z$; from 
these two values we obtain \citep[see Eq. 4 in ][]{RealeProxCen2004} the maximum
temperature of the loop in quiescent conditions (reached at its apex): 
$T_{\rm max,eq}=8.4\times 10^6$\,K and $T_{\rm max,eq}=6.8\times 10^6$\,K,
respectively. These estimates are around the peak temperature of the $EMD$ of
EK Dra and may indicate that the loop where the flare occurred belonged to the
family of loops contributing to the bulk of the X-ray emission of this star.

\bibliographystyle{aa}
\bibliography{biblio}

\end{document}